\documentclass[11pt]{article}
\usepackage{amssymb,latexsym,amsmath}
\usepackage[dvips]{graphicx}
\newtheorem{theo}{Theorem}

\newtheorem{lemm}[theo]{Lemma}

\newtheorem{prop}[theo]{Proposition}
\def\nn{\nonumber}
\def\deg{\mathop{\rm deg}\nolimits}

\def\modulo{\mathop{\rm mod}\nolimits}
\def\qdots{\mathinner{\mkern1mu\raise1pt\vbox{\kern7pt\hbox{.}}\mkern2mu
 \raise4pt\hbox{.}\mkern2mu\raise7pt\hbox{.}\mkern1mu}}

\def\R{{\mathbb R}}

\def\gl{\mathfrak{gl}}
\def\ssl{\mathfrak{sl}}
\def\u{\mathfrak{u}}

\def\osp{\mathfrak{osp}}
\def\lb{[\![}
\def\rb{]\!]}
\newcommand{\q}{\hat q}
\newcommand{\p}{\hat p}
\newcommand{\hQ}{\hat Q}
\newcommand{\hP}{\hat P}
\newcommand{\s}{\hat s}

\def\mybox{\hfill$\Box$}
\begin{document}
\begin{center}
{\Large \bf
Harmonic oscillators coupled by springs: \\[2mm]
discrete solutions as a Wigner Quantum System}\\[5mm]
{\bf S.~Lievens\footnote{E-mail: Stijn.Lievens@UGent.be}, }
{\bf N.I.~Stoilova}\footnote{E-mail: Neli.Stoilova@UGent.be; Permanent address:
Institute for Nuclear Research and Nuclear Energy, Boul.\ Tsarigradsko Chaussee 72,
1784 Sofia, Bulgaria} {\bf and J.\ Van der Jeugt}\footnote{E-mail:
Joris.VanderJeugt@UGent.be}\\[1mm]
Department of Applied Mathematics and Computer Science,
Ghent University,\\
Krijgslaan 281-S9, B-9000 Gent, Belgium.
\end{center}

\vskip 3cm

\begin{abstract}
We consider a quantum system consisting of a one-dimensional chain of $M$ identical 
harmonic oscillators with natural frequency~$\omega$, coupled by means of springs.
Such systems have been studied before, and appear in various models.
In this paper, we approach the system as a Wigner Quantum System, not imposing
the canonical commutation relations, but using instead weaker relations following from the compatibility
of Hamilton's equations and the Heisenberg equations.
In such a setting, the quantum system allows solutions in a finite-dimensional
Hilbert space, with a discrete spectrum for all physical operators.
We show that a class of solutions can be obtained using generators
of the Lie superalgebra $\gl(1|M)$. 
Then we study the properties and spectra of the physical operators 
in a class of unitary representations of $\gl(1|M)$.
These properties are both interesting and intriguing. 
In particular, we can give a complete analysis of the eigenvalues
of the Hamiltonian and of the position and momentum operators (including multiplicities).
We also study probability distributions of position operators when
the quantum system is in a stationary state, and the effect of the position of
one oscillator on the positions of the remaining oscillators in the chain.
\end{abstract}

\vfill\eject

%

\setcounter{equation}{0}
\section{Introduction} \label{sec:Introduction}%

In recent years quantum information theory has known an enormous expansion.
This has boosted new interest in probabilistic and geometric aspects of state spaces
of simple quantum systems.
In this context, the dynamics of entanglement in a chain of coupled harmonic oscillators
has been the subject of many papers~\cite{Audenaert,Brun,Eisert,Halliwell,Plenio}.
One of the systems for which entanglement dynamics is being studied consists of
a large chain of harmonic oscillators coupled by some nearest neighbour interaction~\cite{Plenio}. 
In a popular model this coupling is represented by springs obeying Hooke's law. 
Then the Hamiltonian of the system is given by:
\begin{equation}
\hat{H}=\sum_{k=1}^{M} \Big( \frac{\hat{p}_k^2}{2m}
+ \frac{m\omega^2}{2} \hat{q}_k^2 + \frac{cm}{2}(\hat{q}_k-\hat{q}_{k+1})^2  \Big). 
\label{Intro-H}
\end{equation}
In other words, the quantum system consists of a string or chain of $M$ identical 
harmonic oscillators, each having the same mass $m$ and natural frequency $\omega$.
The position and momentum operator for the $k$th oscillator are given by $\q_k$ and $\p_k$;
more precisely $\q_k$ measures the displacement of the $k$th mass point with respect to its
equilibrium position (see Figure~1).
The last term in~(\ref{Intro-H}) represents the nearest neighbour coupling by
means of ``springs'', with a coupling strength~$c$ ($c>0$). Finally, we shall assume
periodic boundary conditions, i.e.
\begin{equation}
\q_{M+1}\equiv \q_1.
\label{qM+1}
\end{equation}
Such quantum systems are also relevant in quantum optics (photonic crystals), or
for describing phonons in a crystal~\cite{Plenio,Cohen}. 

\begin{figure}[htb]
 \begin{center}
 \includegraphics{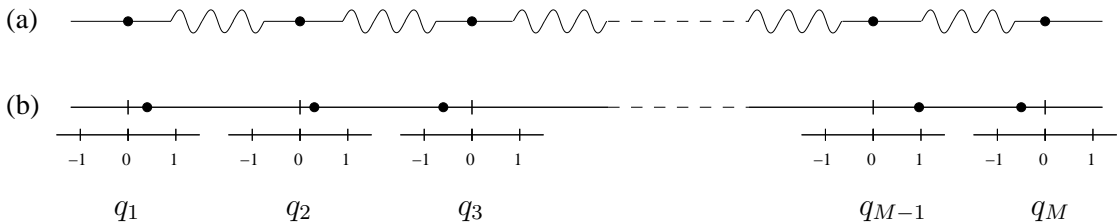}
 \end{center}
\caption{Illustration of the quantum system: (a) the $M$ masses in equilibrium position, (b)
certain displacements $q_k$ for each oscillator.}
\end{figure}

In the standard approach for the quantum system governed by~(\ref{Intro-H}), one
assumes the canonical commutation relations (CCR's)
\begin{equation}
	[\q_k,\q_l]=0, \qquad [\p_k,\p_l]=0,\qquad [\q_k,\p_l]=i\hbar\delta_{kl}.
	\label{CCR}
\end{equation}
Then, reformulating the problem in normal coordinates, the eigenstates of $\hat H$ 
can be described in some infinite-dimensional Fock space~\cite{Plenio}. 

The approach of the present paper is more general. 
Instead of postulating the CCR's, we shall start from a more general quantization procedure.
This procedure is based upon the compatibility of Hamilton's equations with the Heisenberg equations.
Such systems are called Wigner Quantum Systems (WQS's)~\cite{Palev86}.
The idea is based on Wigner's observation -- for the simple example of a one-dimensional 
harmonic oscillator -- that this quantum system also allows solutions for which both
Hamilton's equations and the Heisenberg equations are satisfied, but not the CCR's~\cite{Wigner}. 
In other words, the CCR's are sufficient but not necessary conditions for 
Hamilton's equations and the Heisenberg equations to be compatible.
Wigner's work led to the theory of parabosons and parafermions in 
quantum field theory~\cite{Green,Kamefuchi,Ryan}, 
and because of this attention its impact for ordinary 
quantum systems was somewhat overlooked. 
Another reason why WQS's did not receive immediate attention was because 
no general solutions could be constructed for the compatibility conditions 
of simple WQS's (apart from the one-dimensional harmonic oscillator). 
It was only much later -- after the theory of Lie superalgebra's was completed -- 
that Palev~\cite{Palev1} observed that classes of 
WQS-solutions for the $n$-dimensional harmonic oscillator are described by means 
of representations of the Lie superalgebras $\osp(1|2n)$ and $\ssl(1|n)$ or $\gl(1|n)$. 
This algebraic or representation theoretic approach to quantum systems has revived 
the interest in WQS's~\cite{Degasperis, Atakishiyev, Lohe}.
The WQS approach has so far been applied to simple systems of free
harmonic oscillators, with some interesting and surprising results~\cite{PS1,PS2,KPSV1,KPSV2,Palev2}.
Here it is, for the first time, applied to a more realistic quantum system.

In this paper, we shall study the system described by~(\ref{Intro-H}) as
a WQS. This implies that, apart from the standard solutions for which the CCR's hold,
we shall also discover non-canonical solutions. In particular, we shall
show that the quantum system allows solutions in a finite-dimensional Hilbert space,
thus with a discrete spectrum not only for $\hat H$ but also for the position and
momentum operators $\q_k$ and $\p_k$. 
The class of solutions considered here is related to a $\gl(1|M)$ solution of
the quantization or compatibility conditions, and we consider then a simple
class of $\gl(1|M)$ representations in which the physical properties of the
quantum system are analysed.

In Section~\ref{sec:Q} we analyse the compatibility conditions for the system~(\ref{Intro-H})
as a WQS. This leads to an expression of the Hamiltonian $\hat H$ in terms of
a set of $2M$ operators $a_r^\pm$ ($r=1,\ldots,M$), which are themselves certain
lineair combinations of the $\q_k$ and $\p_k$. These operators $a_r^\pm$ should
satisfy certain triple relations, see~(\ref{algrelations}). The task is then
to construct operator solutions (constructions for the $a_r^\pm$ as operators acting
in some Hilbert space) satisfying these triple relations. 
Although a complete set of solutions cannot be given, we show in Section~\ref{sec:alg}
that certain generators of the Lie superalgebra $\gl(1|M)$ satisfy the triple relations.
In other words, we present algebraic solutions of the compatibility condition (or quantization
condition) in terms of a Lie superalgebra. 
In order to have algebraic solutions that also satisfy the required unitarity conditions,
the coupling constant $c$ is bounded by a critical value $c_0$, depending upon~$M$.
Since the problem has a $\gl(1|M)$ solution, the unitary representations of $\gl(1|M)$
serve as Hilbert space representations for the operators of the quantum system.
In principle, all unitary representations of $\gl(1|M)$ are allowed for this purpose.
In this paper, however, we shall concentrate on a particular class of unitary representations
$W(p)$, mainly for computational purposes. For these representations $W(p)$, the actions
of the $\gl(1|M)$ generators, and in particular of the operators $a_r^\pm$, are very
simple expressions, see Section~\ref{sec:repres}. On the other hand, the class of representations $W(p)$
is already sufficiently rich to exhibit intriguing physical properties of the 
quantum system under consideration.
These properties are examined in the remaining sections.
In Section~\ref{sec:H} we study the energy spectrum of the quantum system in $W(p)$. 
If $c=0$ (absence of coupling), the system consist of $M$ independent (or free) identical one-dimensional
oscillators, and it is easy to see that there are $M+1$ equidistant energy levels in $W(p)$,
with multiplicities $\binom{M}{k}$ according to the level~$k$ ($k=0,1,\ldots,M)$.
For $0<c<c_0$, these energy levels split: we can give a closed formula for the 
energy levels themselves, for their multiplicities, and for the total number of
energy levels (which grows like $3^{M/2}$).
Section~\ref{sec:position} is devoted to the investigation of spatial properties of
the chain of coupled oscillators, if the system is in one of the representations $W(p)$.
Clearly, since the representations considered here are finite-dimensional,
the spectrum of position operators $\q_k$ and momentum operators $\p_k$ is discrete.
We manage to give the spectrum of these operators in closed form, but so far an analytic expression
for a set of orthonormal eigenvectors is missing.
Finally, we examine numerically for a simple example ($M=4$) some position 
probability distributions of the oscillator system.
The case of {\em atypical} representations $W(p)$ (i.e.\ $p\leq M-1$) is examined
in Section~7, followed by some concluding remarks in Section~8.

\setcounter{equation}{0}
\section{The quantization procedure} \label{sec:Q}

In our approach we shall require that Hamilton's equations
\begin{equation}
    {\dot{\hat{q}}}_k=\frac{\partial \hat H}{\partial \hat{p}_k}, \quad 
		{\dot{\hat{p}}}_k=-\frac{\partial \hat H}{\partial \hat{q}_k} \qquad (k=1,2,\ldots,M)
     \label{Ham}
\end{equation}
and the Heisenberg equations
\begin{equation}
     {\dot{\hat{p}}}_k = \frac{i}{\hbar}[\hat{H},\hat{p}_k], \quad
     {\dot{\hat{q}}}_k = \frac{i}{\hbar}[\hat{H},\hat{q}_k]  \qquad (k=1,2,\ldots,M)
     \label{Heis}
\end{equation}
should be identical as operator equations.
Since Hamilton's equations for the Hamiltonian~(\ref{Intro-H}) take the explicit form
\begin{align}
	\dot{\q}_k &= \frac{1}{m} \p_k, \label{qdot}\\
	\dot{\p}_k &= cm\,\q_{k-1} -m(\omega^2+2c)\,\q_k+cm \,\q_{k+1}, \label{pdot}
\end{align}
the compatibility conditions read
\begin{align}
  [\hat{H},\hat{q}_k] &= -\frac{i\hbar}{ m} \hat{p}_k , \label{CCpq1}\\{} 
  [\hat{H},\hat{p}_k] &=-i\hbar cm\,\q_{k-1}+i\hbar m(\omega^2+2c)\,\q_k-i\hbar cm \,\q_{k+1}, \label{CCpq2}
\end{align}
where $k=1,2,\ldots,M$, and -- extending~(\ref{qM+1}) -- $\q_0$ stands for $\q_M$, or
more generally
\begin{equation}
\q_k = \q_{\;k\modulo M}, \quad \p_k = \p_{\;k\modulo M}
\label{modulo-qp}
\end{equation}
whenever $k$ is out of the range $\{1,2,\ldots,M\}$. 
In other words, the task is to find operator solutions for $\q_k$ and $\p_k$ such that 
the compatibility conditions~(\ref{CCpq1})-(\ref{CCpq2}), together with~(\ref{Intro-H}),
are satisfied. Furthermore, since $\q_k$ and $\p_k$ correspond to physical observables,
the operators should be unitary:
\begin{equation}
\q_k^\dagger = \q_k, \quad \p_k^\dagger = \p_k \qquad (k=1,2,\ldots,M).
\label{unitaryqp}
\end{equation}

Just as in the canonical treatment of the problem~\cite{Plenio,Cohen}, 
it will be useful to introduce ``normal
coordinates'' and reformulate the problem in terms of these new coordinates.
So let us consider the finite Fourier transforms of $\q_k$ and $\p_k$:
\begin{align}
\hat{Q}_r &= \frac{1}{\sqrt{M}}\sum_{k=1}^{M} e^{-\frac{2\pi irk}{M}} \hat{q}_k,\label{Qr}\\
\hat{P}_r &= \frac{1}{\sqrt{M}}\sum_{k=1}^{M} e^{\frac{2\pi irk}{M}} \hat{p}_k. \label{Pr}
\end{align}
The inverse relations are given by:
\begin{align}
\hat{q}_k &= \frac{1}{\sqrt{M}}\sum_{r=1}^{M} e^{\frac{2\pi ikr}{M}} \hat{Q}_r, \label{qk}\\
\hat{p}_k &= \frac{1}{\sqrt{M}}\sum_{r=1}^{M} e^{-\frac{2\pi ikr}{M}} \hat{P}_r. \label{pk}
\end{align}
As for $\q_k$ and $\p_k$, see~(\ref{modulo-qp}), it will sometimes be useful to extend the indices by
\begin{equation}
\hQ_k = \hQ_{\;k\modulo M}, \quad \hP_k = \hP_{\;k\modulo M}.
\label{modulo-QP}
\end{equation}
Note that the unitarity conditions~(\ref{unitaryqp}) imply:
\begin{equation}
\hQ_r^\dagger = \hQ_{M-r}, \quad \hP_r^\dagger = \hP_{M-r} \qquad (r=1,2,\ldots,M),
\label{unitaryQP}
\end{equation}
and in particular, following the convention~(\ref{modulo-QP}), 
$\hQ_M^\dagger = \hQ_{M}$ and $\hP_M^\dagger = \hP_{M}$.
An essential part is now to substitute~(\ref{qk})-(\ref{pk}) in the Hamiltonian $\hat H$, 
given by~(\ref{Intro-H}), and simplify this expression without assuming any
commutation relations between the operators. 
For the term $\sum_k \p_k^2 = \sum_k \p_k \p_k^\dagger$, one can, after the substitution,
use the identity
\[
\sum_{k=1}^M e^{\frac{2\pi ik(r-s)}{M}}=M\delta_{rs},
\]
and thus $\sum_k \p_k^2 = \sum_r \hP_r \hP_r^\dagger$. Similarly, 
$\sum_k \q_k^2 = \sum_r \hQ_r \hQ_r^\dagger$. The coupling terms can be written as
\begin{equation}
\sum_k (\q_k-\q_{k+1})(\q_k-\q_{k+1})^\dagger =
\sum_k \q_k\q_k^\dagger + \sum_k \q_{k+1}\q_{k+1}^\dagger -
\sum_k (\q_k\q_{k+1}^\dagger+ \q_{k+1}\q_{k}^\dagger).
\end{equation}
In the right hand side, the first two sums are of the same form as before; the last sum yields
\begin{align}
	\sum_k (\q_k\q_{k+1}^\dagger+ \q_{k+1}\q_{k}^\dagger) &=
	  \sum_r (e^{\frac{-2\pi i r}{M}}\hQ_r\hQ_r^\dagger+ e^{\frac{2\pi i r}{M}}\hQ_{r}\hQ_r^\dagger)\nn\\
	  &= \sum_r 2\cos(\frac{2\pi r}{M}) \hQ_r\hQ_r^\dagger.
\end{align}
Thus we obtain, just as in the canonical case~\cite{Plenio,Cohen}
\begin{equation}
\hat{H}=\sum_{r=1}^{M} \Big( \frac{1}{2m}\hat{P}_r\hat{P}_r^\dagger
+ \frac{m\omega_r^2}{2} \hat{Q}_r\hat{Q}_r^\dagger \Big), 
\label{HPQ}
\end{equation}
where, for $r=1,2,\ldots,M$, the quantities $\omega_r$ are positive numbers with 
\begin{equation}
\omega_r^2=\omega^2 +2c-2c\cos(\frac{2\pi r}{M})=\omega^2+4c\sin^2(\frac{\pi r}{M}),
\label{omega}
\end{equation}
and clearly
\begin{equation}
\omega_{M-r}=\omega_r. 
\label{symm-omega}
\end{equation}
Substituting~(\ref{qk})-(\ref{pk}) in~(\ref{CCpq1})-(\ref{CCpq2}) yields the set of
compatibility conditions for the new operators:
\begin{align}
[\hat{H},\hat{Q}_r]&=-\frac{i\hbar}{ m} \hat{P}_r^\dagger ,\quad
   [\hat{H},\hat{Q}_r^\dagger ]=-\frac{i\hbar}{ m} \hat{P}_r, \label{CCPQ1} \\{} 
[\hat{H},\hat{P}_r] &= i\hbar m\omega_r^2\hat{Q}_r^\dagger, \quad 
[\hat{H},\hat{P}_r^\dagger ]= i\hbar m\omega_r^2\hat{Q}_r. \label{CCPQ2}
\end{align}
The task is now reduced to finding operator solutions for $\hQ_r$ and $\hP_r$ such that 
the compatibility conditions~(\ref{CCPQ1})-(\ref{CCPQ2}), together with~(\ref{HPQ}),
are satisfied.

As a final step it is convenient to introduce linear combinations of the unknown
operators $\hQ_r$ and $\hP_r$, say $a_r^+$ and $a_r^-$ ($r=1,2,\ldots,M$), by
\begin{align}
a_{r}^- &= \sqrt{\frac{m \omega_r}{2\hbar}}
 \hat{Q}_{r} + \frac{i} {\sqrt {2m \omega_r \hbar}} \hat{P}_{r}^\dagger, \label{a-}\\
a_{r}^+ &= \sqrt{\frac{m \omega_r}{2\hbar}}
 \hat{Q}_{r}^\dagger - \frac{i}{\sqrt {2m \omega_r \hbar}} \hat{P}_{r},  \label{a+}
\end{align} 
with
\begin{equation}
(a_r^\pm)^\dagger = a_r^\mp. \label{unitarya}
\end{equation}
Observe that the inverse relations take the form
\begin{align}
\hat{Q}_r &= \sqrt{\frac{\hbar}{2m\omega_r}} \big( a_{M-r}^++a_r^-\big), \;\;r=1,\ldots,M-1; \;\;
\hat{Q}_M=\sqrt{\frac{\hbar}{2m\omega_M}} \big( a_{M}^++a_M^-\big); \label{Qa}\\
\hat{P}_r &= i\sqrt{\frac{m\omega_r\hbar}{2}} \big( a_{r}^+-a_{M-r}^-\big), \;\;r=1,\ldots,M-1; \;\;
\hat{P}_M=i\sqrt{\frac{m\omega_M\hbar}{2}} \big( a_{M}^+-a_M^-\big); \label{Pa}
\end{align}
with similar expressions for $\hQ_r^\dagger$ and $\hP_r^\dagger$.
In terms of the new set of unknown operators $a_r^\pm$ ($r=1,2,\ldots,M$), the Hamiltonian~(\ref{HPQ})
becomes:
\begin{equation}
\hat{H}=\sum_{r=1}^{M}  \frac{\hbar \omega_r}{2} \{ a_r^-, a_r^+\}= 
\sum_{r=1}^{M}  \frac{\hbar \omega_r}{2} (a_r^- a_r^+ + a_r^+ a_r^-). 
\label{Haa}
\end{equation}
It is essential -- and the reader should verify this -- that in going from~(\ref{HPQ}) to~(\ref{Haa}),
no commutation relations among the operators $\hQ_r$ and $\hP_r$ are used, but only 
identities like~(\ref{modulo-QP}), (\ref{unitaryQP}) and~(\ref{symm-omega}).
A final and simple calculation, using~(\ref{a-}) and~(\ref{a+}), shows that~(\ref{CCPQ1})-(\ref{CCPQ2})
is equivalent to 
\begin{equation}
[\hat{H},a_r^\pm ]=\pm \hbar \omega_r a_r^\pm, \quad \ (r=1,2\ldots,M). 
\label{CCa}
\end{equation}
Thus
\begin{prop}
In the approach of~(\ref{Intro-H}) as a Wigner Quantum System, the problem is reduced
to finding $2M$ operators $a_r^\pm$ ($r=1,\ldots,M)$, acting in some Hilbert space,
such that $(a_r^\pm)^\dagger = a_r^\mp$ and
\begin{equation}
[\sum_{j=1}^{M}  \omega_j (a_j^- a_j^+ + a_j^+ a_j^-) , a_r^\pm ]=
\pm 2 \omega_r a_r^\pm, \quad \ (r=1,2\ldots,M). 
\label{algrelations}
\end{equation}
The operators corresponding to physical observables $\q_k$ and $\p_k$ are then known linear
combinations of $a_r^\pm$, and the Hamiltonian $\hat H$ is given by~(\ref{Haa}).
\end{prop}
As we shall see in the following section, this is an algebraic problem that has a
class of solutions in terms of the Lie superalgebra $\gl(1|M)$.

Before concentrating on an algebraic solution, let us end this section with a few words
about the time dependency of the operators.
The time dependency of $\q_k$ and $\p_k$ is determined by~(\ref{qdot})-(\ref{pdot}). From these
equations and~(\ref{Qr})-(\ref{Pr}) it follows that
\begin{equation}
\dot \hQ_r = \frac{1}{m} \hP_r^\dagger,\quad \dot \hP_r = -m\omega_r^2\hQ_r^\dagger.
\label{QPdot}
\end{equation}
Using (\ref{a-})-(\ref{a+}) yields
\begin{equation}
\dot a_r^- = -i\omega_r a_r^-, \quad \dot a_r^+ = i\omega_r a_r^+,
\label{adot}
\end{equation}
with the evident solution
\begin{equation}
a_r^\pm (t) = e^{\pm i\omega_r t} a_r^\pm (0).
\label{a(t)}
\end{equation}
So it is sufficient to have solutions for the operators $a_r^\pm$ at time 0, $a_r^\pm \equiv a_r^\pm(0)$.
The time dependence for $a_r^\pm(t)$ is given by~(\ref{a(t)}), and since all operators of the quantum
system can be expressed in terms of $a_r^\pm$, their time dependence follows.
As a consequence, we shall concentrate on solutions of the system~(\ref{Haa})-(\ref{CCa}) at time $t=0$.

For completeness, it should also be mentioned that if the CCR's~(\ref{CCR}) hold, then 
the operators $a_r^\pm$ satisfy the usual boson relations $[a_r^\pm,a_s^\pm]=0$, $[a_r^-,a_s^+]=\delta_{rs}$.
In that case~(\ref{CCa}) follows automatically from~(\ref{Haa}). 

\setcounter{equation}{0}
\section{Algebraic solutions of the compatibility conditions} \label{sec:alg}

The set of relations~(\ref{algrelations}), together with the conditions~(\ref{unitarya}),
are reminiscent of the algebraic relations satisfied by a set of $\gl(1|M)$ generators~\cite{Palev1}. 
We shall show that our problem has indeed a solution in terms of the Lie superalgebra $\gl(1|M)$ or $\ssl(1|M)$.
Let us first recall the definition of $\gl(1|M)$: it is a Lie superalgebra with basis
elements $e_{jk}$, with $j,k=0,1,\ldots,M$. The elements $e_{k0}$ and $e_{0k}$ ($k=1,\ldots,M$)
are {\em odd} elements, having degree $\deg(e_{k0})=\deg(e_{0k})=1$; the remaining basis elements 
are {\em even} elements, having degree~0. The Lie superalgebra bracket 
is determined by~\cite{Kac1,Kac2,Scheunert}
\begin{equation}
\lb e_{ij}, e_{kl} \rb = \delta_{jk} e_{il} - (-1)^{\deg(e_{ij})\deg(e_{kl})} \delta_{il}e_{kj}.
\label{eij}
\end{equation}
In a representation, or in the enveloping algebra of $\gl(1|M)$, the bracket $\lb x,y\rb$ 
(where $x$ and $y$ are homogeneous elements of $\gl(1|M)$) stands
for an anti-commutator if $x$ and $y$ are both odd elements, and for a commutator otherwise.
The Lie superalgebra $\ssl(1|M)$ is the subalgebra of $\gl(1|M)$ consisting of elements with supertrace~0,
or also $\ssl(1|M)= \lb \gl(1|M), \gl(1|M) \rb$.

For a Lie superalgebra one can also fix a star condition, i.e.\ an anti-linear anti-involution.
For $\gl(1|M)$ or $\ssl(1|M)$ such a star condition is fixed by a signature $\sigma$, i.e.\
a sequence of plus or minus signs $\sigma=(\sigma_1,\ldots,\sigma_M)$ and
\begin{equation}
(e_{0k})^\dagger = \sigma_k e_{k0},\quad (k=1,\ldots,M)
\label{starcondition}
\end{equation}
thus with each $\sigma_k$ equal to $+1$ or $-1$. 
We are particularly interested in the case where all $\sigma_k$'s are $+1$ since this 
corresponds to the ``compact form'' $\u(1|M)$ of $\gl(1|M)$~\cite{Parker}, 
for which finite-dimensional unitary representations exist~\cite{Gould}.

We shall now show that there exist solutions of the form
\begin{equation}
a_k^- = \sqrt{\frac{2}{\omega_k}}\; \alpha_k\, e_{k0}, \quad
a_k^+ = \sqrt{\frac{2}{\omega_k}}\; \alpha_k^*\, \sigma_k\, e_{0k}, \quad(k=1,\ldots,M)
\label{a=e}
\end{equation}
with $\alpha_k$ certain complex constants to be determined.
First of all, note that by~(\ref{starcondition}), the unitarity condition~(\ref{unitarya})
is automatically satisfied. 
With~(\ref{a=e}), the Hamiltonian~(\ref{Haa}) becomes
\begin{equation}
\hat H = \hbar\Bigl(\sum_{j=1}^M \sigma_j |\alpha_j|^2\Bigr)\; e_{00} + 
\hbar \sum_{k=1}^M \sigma_k |\alpha_k|^2\; e_{kk},
\label{Hee}
\end{equation}
and the commutator of the above with~(\ref{a=e}) yields:
\begin{equation}
[\hat H, a_k^\pm ] = \pm \hbar \Bigl(\sum_{j=1}^M \sigma_j |\alpha_j|^2 - \sigma_k |\alpha_k|^2\Bigr) a_k^\pm \qquad
(k=1,\ldots,M).
\end{equation}
These should coincide with the compatibility conditions~(\ref{CCa}).
Thus we get a system of $M$ equations in the unknown coefficients $\alpha_k$:
\begin{equation}
\sum_{j=1}^M \sigma_j |\alpha_j|^2 - \sigma_k |\alpha_k|^2 = \omega_k\qquad (k=1,2,\ldots,M).
\label{eqs-alpha}
\end{equation}
It is easy to verify that a solution for this set of equations is determined by
\begin{equation}
\sigma_k |\alpha_k|^2 = -\omega_k + \frac{1}{M-1}\sum_{j=1}^M \omega_j.
\label{alpha}
\end{equation}
For further use, it will be convenient to introduce the following numbers:
\begin{equation}
\beta_k = -\omega_k + \frac{1}{M-1}\sum_{j=1}^M \omega_j,
\label{beta}
\end{equation}
with $\omega_j$ the values fixed by~(\ref{omega}). 
Note that 
\begin{equation}
\beta_{M-k}=\beta_k, \qquad \beta \equiv \sum_{j=1}^M \beta_j = \sum_{j=1}^M \omega_j,
\label{symm-beta}
\end{equation}
and thus $\hat H$ can be rewritten as
\begin{equation}
\hat H = \hbar(\beta\, e_{00} + \sum_{k=1}^M \beta_k\, e_{kk}).
\label{Hebeta}
\end{equation}

Remember that we are primarily interested in the signature with all $\sigma_k$'s equal to $+1$.
Since $\sigma_k |\alpha_k|^2 = \beta_k$, the question is whether there exist solutions such
that all $\beta_k$'s are positive.
At first sight, (\ref{beta}) indicates that $\beta_k$ is equal to $-\omega_k$ plus ``some average
value'' of the $\omega_j$'s, and hence one would expect half of the $\beta_k$'s to be negative and 
half of them to be positive. 
We shall show, however, that under certain conditions (``weak coupling'', i.e.\ a small value for $c$),
all $\beta_k$'s are indeed positive. 
First of all, note that for $c>0$,
\begin{equation}
\beta_1 > \beta_2 > \cdots > \beta_{\lfloor M/2\rfloor}, \quad \beta_{\lfloor M/2\rfloor} \leq
\beta_{\lfloor M/2\rfloor+1 }< \cdots < \beta_M,
\end{equation}
since a similar property holds for the values $\omega_k$. 
Thus if $\beta_{\lfloor M/2\rfloor}>0$, then all $\beta_k$'s are positive. 
The value of $\beta_{\lfloor M/2\rfloor}$ depends on the value of $c$; if $c=0$ then indeed all 
$\beta_k=\omega/(M-1)$ are positive. So by continuity as a function of~$c$
there will be a certain interval $]0,c_0[$ where all
$\beta_k$'s are positive. This critical value $c_0$ is the $c$-value for which 
$\beta_{\lfloor M/2\rfloor}=0$. For general $M$, this is a complicated transcendental equation that can
be solved only numerically. Table~1 gives the numerical solutions for this 
equation, for $M$ ranging from 4 to 21 (for $M=2$ or $M=3$ the
$\beta_k$'s are always positive).
\begin{table}[htb]
\caption{Critical values $c_0/\omega^2$}
\begin{center}
\begin{tabular}{||r|r||r|r||}
\hline
$M$ & $c_0/\omega^2$ & $M$ & $c_0/\omega^2$ \\
\hline
4 & 0.9873724357 & 13 & 0.10546881460 \\
5 & 0.7500000000 & 14 & 0.09256321610 \\
6 & 0.3457442295 & 15 & 0.08687882025 \\
7 & 0.2982653656 & 16 & 0.07814800074 \\
8 & 0.2061705212 & 17 & 0.07388896853 \\
9 & 0.1851128402 & 18 & 0.06760983697 \\
10 & 0.1464642846 & 19 & 0.06429500840 \\
11 & 0.1343028683 & 20 & 0.05957194222 \\
12 & 0.1134651313 & 21 & 0.05691629341 \\
\hline
\end{tabular}
\end{center}
\end{table}
The following proposition gives a lower bound for the critical value $c_0$, such
that all $\beta_k$'s are positive.
\begin{prop}
An upper bound for $c$ is determined by:
\begin{equation}
0\leq c \leq \frac{\omega^2}{2(M-2)} \implies \beta_k \geq 0,\quad 
\text{for}\  1\leq k\leq M.
\label{c-bound}
\end{equation}
\end{prop}

\noindent {\bf Proof.}
By definition~(\ref{beta}), $\beta_k\geq 0$ if and only if
\begin{equation}
\omega_k \leq \frac{1}{M-1}\sum_{j=1}^M \omega_j 
 \iff (M-1)^2\omega_k^2 \leq  \sum_{j=1}^M \omega_j^2 + 
2\sum_{i>j} \omega_i \omega_j.
\label{inequality}
\end{equation}
We write the right hand side of~(\ref{inequality}) as a series with respect to~$c$.
For the first sum, we find:
\begin{equation*}
\sum_{j=1}^M \omega_j^2  = \sum_{j=1}^M \left(\omega^2 + 4c\sin^2(\frac{j\pi}{M}) \right) 
 = M\omega^2 + 2c\sum_{j=1}^M \left( 1-\cos(\frac{2j\pi}{M}) \right) 
 = M(\omega^2 + 2c).
\end{equation*}
Here, we used the Lagrange identity:
\begin{equation*}
1 + \cos(x) + \cos(2x) + \cdots + \cos(nx) 
= \frac{1}{2}+\frac{\sin(\frac{(2n+1)x}{2})}{2\sin(\frac{x}{2})}.
\end{equation*}
In order to evaluate the second sum in~(\ref{inequality}), let
\begin{equation}
	F(c) \equiv \sum_{i>j} \omega_i \omega_j = 
	\sum_{i>j} \sqrt{\omega^2 + 4c\sin^2(\frac{i\pi}{M})}  
	\sqrt{\omega^2 + 4c\sin^2(\frac{j\pi}{M})}.
\label{F}	
\end{equation}
It is then clear that
\begin{equation*}
F(0) = \omega^2 \frac{M(M-1)}{2}.
\end{equation*}
One can write the derivate of $\omega_k$ with respect to~$c$ as:
\begin{equation*}
\omega_k' = \frac{d \omega_k}{d c} = \frac{2}{\omega_k}\sin^2(\frac{k\pi}{M}),
\end{equation*}
and one finds that
\begin{equation*}
F'(c) = \sum_{j>i} (\omega_i'\omega_j + \omega_i\omega_j')
\geq 0,
\end{equation*}
since $\omega_k' \geq 0$. This implies that $F(0) \leq F(c)$ for $c \geq 0$.
Thus it follows from~(\ref{inequality}) that  
\begin{equation}
(M-1)^2\omega_k^2 \leq \sum_{j=1}^M \omega_j^2  + 2\, F(0) 
\label{suff-cond}
\end{equation}
is a sufficient condition for $\beta_k\geq 0$.
Since $\omega_k^2 \leq \omega_{\lfloor M/2\rfloor}^2 \leq \omega^2 + 4c$ for $1\leq k\leq M$,
it is sufficient to solve the following inequality
\begin{equation*}
(M-1)^2(\omega^2+4c) \leq M(\omega^2 + 2c) + \omega^2 M(M-1)
\iff c \leq \frac{\omega^2}{2(M-2)},
\end{equation*}
leading to~(\ref{c-bound}). \mybox

Note that for $M=2$ or $M=3$ there are no conditions: for $M=2$, $\beta_1=\omega_2>0$ and 
$\beta_2=\omega_1>0$; for $M=3$, $\beta_1=\beta_2=\omega/2>0$ and $\beta_3=(\omega^2+3c)^{1/2}-\omega/2 >0$.

We can now summarize the main result of this section in the following
\begin{prop}
For fixed $M$, let $c$ satisfy
\[
c \leq \frac{\omega^2}{2(M-2)} (\leq c_0)
\]
(no condition if $M=2$ or $M=3$). 
Then the compatibility conditions~(\ref{Haa})-(\ref{CCa}) have a solution for the
operators $a_k^\pm$ in terms of $\gl(1|M)$ generators:
\begin{equation}
a_k^- = \sqrt{\frac{2\beta_k}{\omega_k}}\;  e_{k0}, \quad
a_k^+ = \sqrt{\frac{2\beta_k}{\omega_k}}\;  e_{0k}, \quad(k=1,\ldots,M)
\label{solution}
\end{equation}
with $\beta_k$ given by~(\ref{beta}). The unitarity conditions~(\ref{unitarya})
are equivalent with the star condition 
\begin{equation}
(e_{0k})^\dagger = e_{k0}.
\label{edagger}
\end{equation}
\end{prop}

\setcounter{equation}{0}
\section{A class of $\gl(1|M)$ representations} \label{sec:repres}

In order to study properties of the given quantum system related to the $\gl(1|M)$
solution of the previous section, one should consider representations of $\gl(1|M)$ for which~(\ref{edagger}) holds.
These are known as the unitary representations (or star representations), and have been classified
by Gould and Zhang~\cite{Gould}. 
For the explicit actions of the $\gl(1|M)$ generators on a Gel'fand-Zetlin basis for
these unitary representations, see~\cite{KSV}. 
These actions becomes fairly complicated, so in this paper we will concentrate on a
particular class of representations, the so-called Fock type representations $W(p)$.

Without going into the details of the construction of such representations, 
we briefly summarize their main features here.
Further details can be found in~\cite{Palev0,SV1}

The representations $W(p)$ are labelled by a number $p$, with
either $p\in\{0,1,2,\ldots,M-1\}$ or else $p\in\R$ with $p>M-1$. We describe 
the representation by giving the basis vectors of the representation space~$W(p)$
and the action of the $\gl(1|M)$ generators on these basis vectors.

For $p\in\{0,1,2,\ldots,M-1\}$, the basis vectors of $W(p)$ are given by:
\begin{equation}
w(\theta)\equiv w(\theta_1,\theta_2, \ldots,\theta_M),\qquad
\theta_i\in\{0,1\},\hbox{ and } |\theta|=\theta_1+\cdots+\theta_M \leq p.
\end{equation}
Note that in this case the dimension of $W(p)$ is given by
\begin{equation}
\dim W(p) = \sum_{k=0}^p \binom{M}{k}.
\end{equation}

For $p$ real and $p>M-1$, the basis vectors of $W(p)$ are all vectors 
$w(\theta)\equiv w(\theta_1,\theta_2, \ldots,\theta_M)$ with each $\theta_i\in\{0,1\}$.
Clearly, for $p>M-1$, the dimension of $W(p)$ is given by $2^M$. 

The action of the $\gl(1|M)$ generators on the basis vectors of $W(p)$ is now given by:
\begin{align}
e_{00} w(\theta) &= (p-|\theta|)\ w(\theta); \label{e00} \\
e_{kk} w(\theta) &= \theta_k\ w(\theta),\qquad (1\leq k\leq M); \label{ekk} \\
e_{k0} w(\theta) &= (1-\theta_k) (-1)^{\theta_1+\cdots+\theta_{k-1}} \sqrt{p-|\theta|}\
 w(\theta_1,\ldots, \theta_k+1,\ldots, \theta_M), \quad (1\leq k\leq M); \label{ek0} \\
e_{0k} w(\theta) &= \theta_k (-1)^{\theta_1+\cdots+\theta_{k-1}} \sqrt{p-|\theta|+1}\
 w(\theta_1,\ldots, \theta_k-1,\ldots, \theta_M), \quad (1\leq k\leq M). \label{e0k}
\end{align}
The action of the remaining $\gl(1|M)$ basis elements can be determined from the above formulas,
and one finds (for $1\leq j< k\leq M$):
\begin{align}
e_{jk} w(\theta) &=\theta_k(1-\theta_j) (-1)^{\theta_j+\cdots+\theta_{k-1}}\
  w(\theta_1,\ldots,\theta_j+1,\ldots, \theta_k-1,\ldots,\theta_{M}),  \label{ejk} \\
e_{kj} w(\theta) &= -\theta_j(1-\theta_k) (-1)^{\theta_j+\cdots+\theta_{k-1}}\
  w(\theta_1,\ldots,\theta_j-1,\ldots,\theta_k+1 ,\ldots,\theta_{M}). \label{ekj}
\end{align}

With respect to the inner product 
\begin{equation}
\langle w(\theta) | w(\theta') \rangle = \delta_{\theta,\theta'},
\label{norm}
\end{equation}
the representation $W(p)$ is unitary for the star condition
\[
e_{jk}^\dagger = e_{kj}.
\]

The representations $W(p)$ with $p>M-1$ are {\em typical} irreducible $\gl(1|M)$ representations; those with
$p\in\{0,1,2,\ldots,M-1\}$ are {\em atypical} irreducible $\gl(1|M)$ representations~\cite{Kac2,JHKT}.
In Sections~5 and~6 we shall develop results for the typical representations; the atypical case
is treated in Section~7.

Note, by~(\ref{solution}) and~(\ref{ek0})-(\ref{e0k}) the action of the operators $a_k^\pm$
on the basis vectors $w(\theta)$ of $W(p)$ ($k=1,\ldots,M$) is given by:
\begin{align}
a_k^- w(\theta) &= \sqrt{\frac{2\beta_k}{\omega_k}}(1-\theta_k) (-1)^{\theta_1+\cdots+\theta_{k-1}} \sqrt{p-|\theta|}\
 w(\theta_1,\ldots, \theta_k+1,\ldots, \theta_M),  \label{actiona-} \\
a_k^+ w(\theta) &= \sqrt{\frac{2\beta_k}{\omega_k}} \theta_k (-1)^{\theta_1+\cdots+\theta_{k-1}} \sqrt{p-|\theta|+1}\
 w(\theta_1,\ldots, \theta_k-1,\ldots, \theta_M). \label{actiona+}
\end{align}
So the operators $a_k^\pm$ raise or lower $\theta_k$ by one unit (if allowed). 
This means that, for $p>M-1$, the basis vectors $w(\theta)$ of $W(p)$ have a Fock basis construction,
by letting $|0\rangle = w(1,1,\ldots,1)$. Then $a_k^- |0\rangle =0$ and 
\begin{equation}
w(\theta) \sim (a_1^+)^{1-\theta_1} (a_2^+)^{1-\theta_2} \cdots (a_M^+)^{1-\theta_M} |0\rangle.
\end{equation}
This is the reason why $W(p)$ is referred to as a Fock representation.

\setcounter{equation}{0}
\section{On the spectrum of $\hat H$ in the representations $W(p)$} \label{sec:H}

For any $p>M-1$, the representation $W(p)$ is of dimension $2^M$.
Under the solution~(\ref{solution}), the Hamiltonian $\hat H$ takes the form~(\ref{Hebeta}),
\begin{equation*}
\hat H = \hbar(\beta\, e_{00} + \sum_{k=1}^M \beta_k\, e_{kk}).
\end{equation*}
with $\beta=\sum_{k=1}^M \beta_k = \sum_{k=1}^M \omega_k$. 
Since the actions of $e_{kk}$ ($k=0,1,\ldots,M$) are diagonal in the basis $w(\theta)$,
see~(\ref{e00})-(\ref{ekk}), it follows that the vectors $w(\theta)$ are eigenvectors
for $\hat H$:
\begin{equation}
\hat H \, w(\theta) = \hbar E_\theta \, w(\theta),
\label{Heigenvectors}
\end{equation}
with eigenvalues
\begin{equation}
E_\theta = \beta(p-|\theta|) + \sum_{k=1}^M \theta_k \beta_k =
\beta(p-\frac{M-2}{M-1}|\theta|) - \sum_{k=1}^M \theta_k \omega_k.
\label{Etheta}
\end{equation}
In this expression, $\theta=(\theta_1,\ldots,\theta_M)$, with each $\theta_k\in\{0,1\}$, and
$|\theta|=\sum_{k=1}^M \theta_k$.

In the case without coupling ($c=0$), all $\beta_k$'s are the same:
$\beta_k = \frac{\omega}{M-1}$ and $\beta = \frac{M}{M-1}\omega$. The eigenvalues of $\hat H$ are then
\begin{equation*}
\hbar\omega (p\frac{M}{M-1}-|\theta|).  
\end{equation*}
The multiplicity 
of this eigenvalue is $\binom{M}{|\theta|}$. 
In other words, there are $M+1$ distinct energy levels, equally spaced with steps of unit $\hbar\omega$.
The lowest energy level corresponds to $E_{(1,\ldots,1)} = \omega p \frac{M}{M-1}-\omega M$,
and the highest to $E_{(0,\ldots,0)} = \omega p \frac{M}{M-1}$.

We are mainly interested in the weak coupling case ($c>0$, but $c<c_0$).
Also in this case, it is easy to describe the energy levels through~(\ref{Etheta}),
but the analysis of their multiplicity requires some further attention.
For this purpose, observe that by~(\ref{symm-beta}) 
\begin{equation}
\sum_{k=1}^M \theta_k\beta_k = \sum_{k=1}^M \theta_k\beta_{M-k}
 = \sum_{k=0}^{M-1} \theta_{M-k}\beta_{k} = 
 \sum_{k=1}^{M} \theta_{M-k}\beta_{k},
\label{symm-theta}
\end{equation}
where in the last step we have followed the convention that
$\beta_0 = \beta_M$, and we have set $\theta_0 = \theta_M$.
Let $w(\theta)$ be an arbitrary eigenvector $\hat H$ with eigenvalue $\hbar E_\theta$.  
Obviously, by~(\ref{Etheta}) and~(\ref{symm-theta}) all basis vectors $w(\theta')$
which are obtained by swapping $\theta_i$ and $\theta_{M-i}$ 
for arbitrary indices $i$ yield the same eigenvalue
$\hbar E_\theta$.
The multiplicity of this eigenvalue is thus (at least)
\begin{equation}
2^{\sum_{k=1}^{M}(\theta_k - \theta_{M-k})^2/2}.
\label{multi}
\end{equation}
Further inspection of the last expression in~(\ref{Etheta}) shows that this is indeed
the multiplicity of the eigenvalue, for $0<c<c_0$.

In Figure~2 we give a plot of the energy levels for $M=4$ and $M=5$, as an illustration of the above.
As $c$ increases, the $M+1$ equidistant energy levels for $c=0$ split up in different levels, 
with smaller degeneracies.

\begin{figure}[htb]
\begin{center}
 \begin{tabular}{ccc}
 (a) $M=4$ &\vspace{12mm} & (b) $M=5$ \\
 \includegraphics[height=9cm,width=7cm]{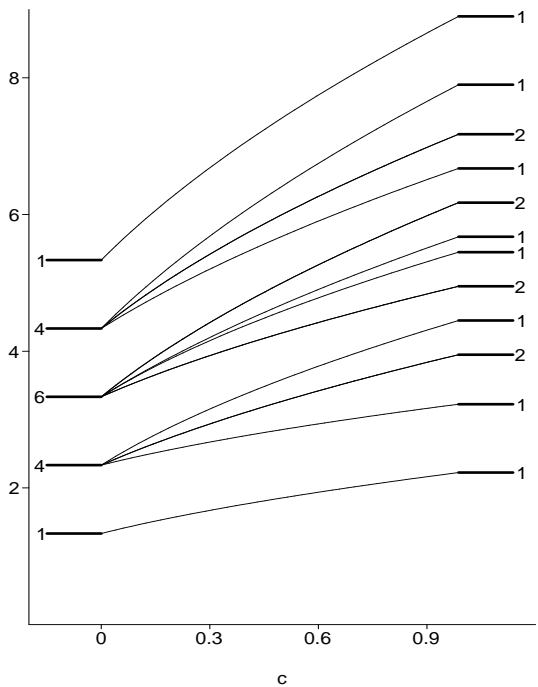} && \includegraphics[height=9cm,width=7cm]{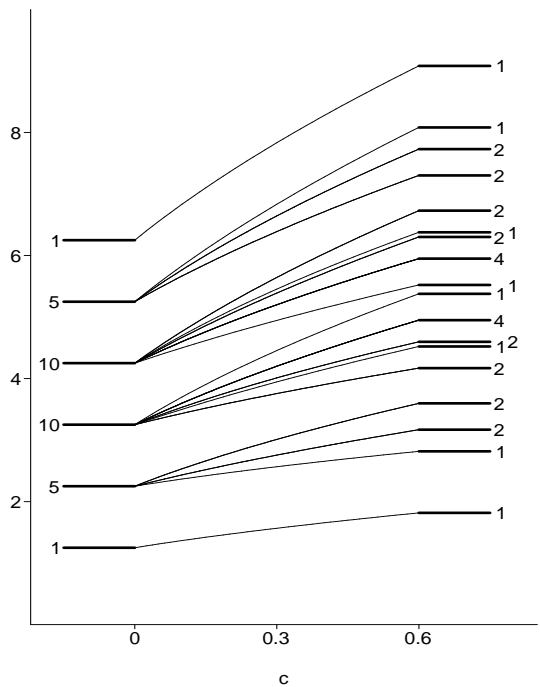}
 \end{tabular}
\end{center} 
\caption{(a) The energy levels of the quantum system for $M=4$ in $W(p)$, 
where we have taken $\hbar=\omega=1$, $p=M$, and $c$ varies from~0 to~$c_0$. 
The vertical axis gives the energy values. The numbers next to the levels
refer to the multiplicity. When $c=0$ there 
are $M+1=5$ energy levels, with multiplicities (1,4,6,4,1). When $0<c<c_0$, the
energy levels split up in 12 levels, with multiplicities 1 or 2.
(b) The same illustration for $M=5$. There are 6 levels for $c=0$ with multiplicities
(1,5,10,10,5,1), and there are 18 levels for $0<c<c_0$, with multiplicities 1, 2 or 4.}
\end{figure}

It is worth noting that one can say something extra about the number of energy levels, for arbitrary~$M$,
also when $c\ne 0$. For $M=4$, the five levels $1+1+1+1+1$ at $c=0$ become $1+3+4+3+1$ levels for $c>0$ [just
counting the energy levels, disregarding their multiplicities], see~Figure~2(a). We list here the
number of levels for a few $M$-values, when $c>0$:
\begin{equation*}
\begin{tabular}{rrc}
$M$ & number of levels & total number of levels\\
1 & 1+1 & 2\\
2 & 1+2+1 & 4\\
3 & 1+2+2+1 & 6 \\
4 & 1+3+4+3+1 & 12\\
5 & 1+3+5+5+3+1 & 18 \\
6 & 1+4+8+10+8+4+1 & 36\\
7 & 1+4+9+13+13+9+4+1 & 54\\
8 & 1+5+13+22+26+22+13+5+1 & 108
\end{tabular}
\end{equation*}
Let $T(M,k)$ ($k=0,1,\ldots,M$) be the number of levels ``per split'' (the numbers in the middle column),
and $L(M)=\sum_{k=0}^M T(M,k)$ be the total number of energy levels for $c>0$.
$T(M,k)$ is also the number of distinct energy levels for all $\theta$ with $|\theta|=k$.
Analysing this, using~(\ref{multi}), one finds
\begin{equation}
T(M,k)=T(M-2,k)+T(M-2,k-1)+T(M-2,k-2).
\end{equation}
Summing now over all $k$ yields a simple recursion for the number of levels $L(M)$:
\begin{equation}
L(M)=3 L(M-2).
\end{equation}
So we obtain the following result for the number of energy levels when $0<c<c_0$, depending on 
whether $M$ is even or odd:
\begin{equation}
L(2n+1)=2\cdot 3^n, \qquad L(2n+2) = 4\cdot 3^{n}, \qquad (n=0,1,2,\ldots).
\end{equation}
Note that the numbers $L(M)$ belong to a known sequence, see entry~A068911
in~\cite{Sloane}, with a simple generating function:
\begin{equation}
G(x)=\sum_{M=0}^\infty L(M)\; x^M = \frac{(1+x)^2}{1-3x^2}.
\end{equation}

\setcounter{equation}{0}
\section{On the spectrum of position operators in the representations $W(p)$
and spatial properties} \label{sec:position}

The purpose of the present section is to consider some geometric aspects of the given
quantum system. In first instance, we shall analyse the spectrum of the position operators $\q_r$
in the representations $W(p)$. By construction, these Hilbert spaces $W(p)$ are finite dimensional,
so the position operators have a discrete spectrum.

The determination of the spectrum for general $M$ and $p$ requires a lot of work. This goes
in two steps: first the operators $\q_r^2$ will be considered; this is the difficult part.
Once the spectrum of $\q_r^2$ is analysed, that of $\q_r$ follows rather easily.
The spectrum of the momentum operators $\p_r^2$ and $\p_r$ is completely similar.
In fact, we shall see that the structure of $\p_r^2$ and $\q_r^2$ as operators in
$\gl(1|M)$ is equivalent, and they can be treated simultaneously.

Note that, due to the symmetry of the system~(\ref{Intro-H}) and~(\ref{qM+1}), the spectrum of $\q_r^2$ and $\q_r$
will be independent of~$r$. 

\subsection{Eigenvalues and eigenvectors for $\q_r^2$}

We start by writing the operator $\q_r^2$ in terms of the $\gl(1|M)$ basis elements.
Using~(\ref{qk}), one finds
\begin{equation*}
\q_r^2 = \frac{1}{2} \{ \q_r, \q_r^\dagger \} = \frac{1}{2M} \sum_{k=1}^M \sum_{l=1}^M 
e^{\frac{2\pi i r(k-l)}{M}} \, \{ \hQ_k, \hQ_l^\dagger \}.
\end{equation*}
Using~(\ref{Qa}) and the solution~(\ref{solution}), this yields
\begin{equation}
\q_r^2 = \frac{\hbar}{mM} \big( \sum_{k=1}^M \frac{\beta_k}{\omega_k^2}\;  e_{00}+
\sum_{k=1}^M \sum_{l=1}^M e^{\frac{2\pi i r(l-k)}{M}}
\frac{\sqrt{\beta_l \beta_k}}{\omega_l\omega_k}\; e_{lk}\bigr).
\label{qr2}
\end{equation}
In a similar way, one finds
\begin{equation}
\p_r^2 =  \frac{\hbar m}{M}   \big( \sum_{k=1}^M \beta_k\;  e_{00}+
\sum_{k=1}^M \sum_{l=1}^M e^{\frac{2\pi i r(l-k)}{M}}\sqrt{\beta_l \beta_k}\; e_{lk}\bigr).
\label{pr2}
\end{equation}
As operators of $\gl(1|M)$, these two expressions are structurally equivalent: $\q_r^2$ is
obtained from the expression of $\p_r^2$ by formally replacing every parameter $\beta_k$ by $\beta_k/\omega_k^2$,
and multiplying by an overall factor $1/m^2$. 
So their spectral analysis is equivalent. 
Since the expression of $\p_r^2$ is somewhat simpler from the point of view of notation (no denominators
in the factors), we shall first concentrate on the analysis for $\p_r^2$.

Following~(\ref{pr2}), it will be useful to introduce the even $\gl(1|M)$ operators
\begin{equation}
\s_r = \sum_{k=1}^M \sum_{j=1}^M e^{\frac{2\pi i r(j-k)}{M}}\sqrt{\beta_j \beta_k} e_{jk}.
\label{sr}
\end{equation}
The main ingredient of our analysis is the following technical lemma:
\begin{lemm}
In the representation $W(p)$, the operator $\s_r$ has only two eigenvalues, namely 
$\beta = \sum_{k=1}^M \beta_k$ and $0$, each with multiplicity $2^{M-1}$. 
\end{lemm}

\noindent {\bf Proof.} 
Consider a vector of the form
\begin{equation}
	v_r \equiv 
	\sum_{k=1}^M e^{\frac{2\pi i r k}{M}} \sqrt{\beta_k} w(1^k),
\label{v_r}	
\end{equation} 
where $w(1^k)$ denotes the basis vector of $W(p)$ with $|\theta| = 1$ and the \lq\lq 1\rq\rq\ 
occurring in position $k$ (counting from left, and starting from one).
For instance, if $M = 4$ then
\begin{equation*}
w(1^1) = w(1,0,0,0),\quad
w(1^2) = w(0,1,0,0),\quad
w(1^3) = w(0,0,1,0),\quad
w(1^4) = w(0,0,0,1).
\end{equation*}
We extend the notation to $w(1^{k_1} \cdots 1^{k_n})$ which denotes the
basis vector with $|\theta| = n $ and with \lq\lq1\rq\rq's in the 
positions $k_1$ up to $k_n$.

It is easy to verify that $v_r$ is an eigenvector of $\hat s_r$ with eigenvalue $\beta$. 

In $\gl(1|M)$, one can see that
\begin{equation}
[e_{l0},\s_r] = -\sqrt{\beta_l} e^{-\frac{2\pi i r l}{M}} 
\sum_{j=1}^M \sqrt{\beta_j} e^{\frac{2\pi i r j}{M}} e_{j0},
\label{el0sr}
\end{equation}
and thus the action~(\ref{ek0}) implies that
\begin{equation}
	[e_{l0},\hat s_r]v_r = 0.
\label{esv}	
\end{equation}
This means that $e_{l0}v_r$ (provided it does not vanish) is also an 
eigenvector of $\s_r$, with the same eigenvalue $\beta$.
Our goal is now to show that
\begin{equation}
\label{eq:v_nr}
e_{l_n0}e_{l_{n-1}0}\cdots e_{l_10}v_r 
\end{equation}
is also an eigenvector of $\s_r$ with the same eigenvalue $\beta$,
provided all ${l_i}$ are different (and different from~$M$).  
If one would apply the same $e_{l0}$ twice, the resulting vector vanishes since
$\{ e_{l0},e_{k0}\} = 0$, by~(\ref{eij}).
This means that we can also assume, without loss of generality,
that $M>l_n > l_{n-1} > \cdots > l_1>0$.

Using~(\ref{ek0}), one can write
\begin{equation}
e_{l_n0}e_{l_{n-1}0}\cdots e_{l_10}v_r  \sim
\sum_{t=0}^n (-1)^t \sum_{k=l_t+1}^{l_{t+1}-1} \sqrt{\beta_k} 
e^{\frac{2\pi i r k}{M}} w(1^{l_1}\cdots 1^{l_t} 1^k 1^{l_{t+1}}\cdots 1^{l_n}),
\label{ev-as-w}
\end{equation}
with $l_0 = 0$ and $l_{n+1} = M+1$.
Using~(\ref{el0sr}), this implies that
\begin{equation}
\label{eq:comm_el0_sn_on_vn}
[e_{l0},\s_r] e_{l_n0}e_{l_{n-1}0}\cdots e_{l_10}v_r \sim
\sum_{j=1}^M \sqrt{\beta_j} e^{\frac{2\pi i r j}{M}}
\sum_{t=0}^n (-1)^t \sum_{k=l_t+1}^{l_{t+1}-1} \sqrt{\beta_k} 
e^{\frac{2\pi i r k}{M}} 
e_{j0} w(1^{l_1}\cdots 1^{l_t} 1^k 1^{l_{t+1}}\cdots 1^{l_n}).
\end{equation}
Keeping in mind the action of $e_{j0}$ on a basis vector $w(\theta)$,
one sees that this expression is a linear combination of basis vectors
$w(\theta)$ for which $|\theta| = n+2$, having a \lq\lq 1\rq\rq\ 
in the positions $l_1$, $l_2$ up to $l_n$ and in two extra positions $x$ and $y$.  
Consider such a vector
\begin{equation*}
w(1^{l_1}\cdots 1^{l_{i_x}} 1^x 1^{l_{i_x+1}}\cdots
1^{l_{i_y}} 1^y 1^{l_{i_y+1}}\cdots 1^{l_n} ).
\end{equation*}
In expression~\eqref{eq:comm_el0_sn_on_vn} this vector will
appear twice, once with $k$ and $j$ playing the role of $x$ and
$y$ respectively, and one vice versa.  In the first case the 
coefficient of this vector is
\begin{equation*}
\sqrt{p - (n+1)} \sqrt{\beta_x\beta_y} e^{\frac{2\pi i r (x+y)}{M}} 
\times (-1)^{i_x}\times(-1)^{i_y+1} ,
\end{equation*}
while in the second case it is
\begin{equation*}
\sqrt{p - (n+1)} \sqrt{\beta_x\beta_y} e^{\frac{2\pi i r (x+y)}{M}} 
\times (-1)^{i_x}\times(-1)^{i_y}.
\end{equation*}
Since these two coefficients cancel, we can conclude that
\begin{equation}
[e_{l0},\s_r] e_{l_n0}e_{l_{n-1}0}\cdots e_{l_10}v_r = 0.
\label{eseeev}
\end{equation}
Following~(\ref{esv}) and~(\ref{eseeev}), all vectors of the form~(\ref{eq:v_nr})
are eigenvectors of $\s_r$ for the eigenvalue~$\beta$. It remains to find the
number of linearly independent eigenvectors, i.e.\ the multiplicity of the eigenvalue $\beta$.

For $n$ fixed, consider the $\binom{M-1}{n}$ vectors~\eqref{eq:v_nr} with
$1\leq l_1 < l_2 < \cdots < l_n \leq M-1$.  
Expressing these in the $w(\theta)$ basis by~(\ref{ev-as-w}),
in total $\binom{M}{n+1}$ basis vectors are involved.
Each basis vector $w(\theta)$ occurs $\binom{n+1}{n} = n+1$ times, \emph{except}
those vectors that have a \lq\lq 1\rq\rq\ in position $M$.  There 
are $\binom{M-1}{n}$ such vectors.  
Next, let $A$ be the $\binom{M-1}{n}\times \binom{M}{n+1}$ 
matrix, consisting of the coefficients of the $\binom{M-1}{n}$ 
vectors~\eqref{eq:v_nr} written in terms of the $\binom{M}{n+1}$ basis vectors.
Select in $A$ those columns that correspond to basis
vector having a \lq\lq1\rq\rq\ in position $M$.  This submatrix is
equivalent with a diagonal matrix of which the diagonal elements 
are proportional to $\sqrt{\beta_M}$.  So, the vectors~\eqref{eq:v_nr} 
with $1\leq l_1 < \cdots <l_n \leq M-1$ are linearly independent.

Furthermore, it is immediately clear that the vectors 
\begin{equation*}
	e_{l_n0}e_{l_{n-1}0}\cdots e_{l_10}v_r \ \text{and}\ 
e_{l_t0}e_{l_{t-1}0}\cdots e_{l_10}v_r
\end{equation*}
are linearly independent if $n\neq t$ (they are linear combinations
of basis vectors with different $|\theta|$).

The conclusion is that we have found
\begin{equation*}
\sum_{n=0}^{M-1} \binom{M-1}{n} = 2^{M-1}
\end{equation*}
linearly independent eigenvectors of $\s_r$ with eigenvalue $\beta$.  
Note that for some fixed $|\theta|$, there
are $\binom{M-1}{|\theta|-1}$ linearly independent eigenvectors
of $\s_r$ with eigenvalue $\beta$.

The eigenvalue $0$ of $\s_r$ also has multiplicity $2^{M-1}$. 
This is seen in a completely similar way, starting with
the vector 
\begin{equation*}
\tilde v_r \equiv 
\sum_{k=1}^M e^{-\frac{2\pi i r k}{M}} (-1)^k \sqrt{\beta_k} w(0^k)
\end{equation*}
and acting repeatedly with $e_{0l}$ (with $1\leq l \leq M-1$) on this vector.
Herein we have extended the notation of~(\ref{v_r}): $w(0^k)$ 
denotes a basis vector where every $\theta_j=1$ ($j\ne k$) except $\theta_k=0$.
For fixed $|\theta|$, there are thus $\binom{M-1}{|\theta|}$ linearly 
independent eigenvectors with eigenvalue $0$.
\mybox

We can now describe the eigenvalues of $\p_r^2$.  Since 
\begin{equation*}
e_{00}\, w(\theta) = (p-|\theta|)\, w(\theta),
\end{equation*}
it follows from~(\ref{pr2}) that an eigenvector of $\s_r$ which is a linear combination 
of basis vectors with fixed $|\theta|$ will also be an eigenvector
of $\p_r^2$.  The eigenvalues of $\p_r^2$ are thus
given by $\frac{\hbar m}{M}(p-K)\beta$, for $0 \leq K \leq M-1$, with
multiplicities $2\binom{M-1}{K}$.  More in particular, the eigenvectors
of $\p_r^2$ with eigenvalue  $\frac{\hbar m}{M}(p-K)\beta$ arise in 
two ways, one set having $|\theta| = K+1$ and containing vectors of 
the form
\begin{equation*}
	e_{l_K0}e_{l_{K-1}0}\cdots e_{l_10}v_r.
\end{equation*}
The other set has $|\theta| = K$ and contains vectors of the form
\begin{equation*}
	e_{0l_{M-K-1}} \cdots e_{0l_1} \tilde v_r. 
\end{equation*}

Unfortunately, the vectors (for a given eigenvalue) constructed here are not orthogonal.
For any fixed $M$, one can construct a set of orthogonal eigenvectors by Gram-Schmidt 
orthogonalisation. But so far we cannot give a closed analytic expression for some
set of orthogonal eigenvectors.

Let us also state the result for the squared position operators $\q_r^2$.
Recall from the beginning of this subsection that every $\beta_k$ should
be replaced by $\beta_k/\omega_k^2$, thus $\beta$ should be replaced by
\begin{equation}
\gamma=\sum_{k=1}^M \frac{\beta_k}{\omega_k^2}.
\label{gamma}
\end{equation}

\begin{prop}
The operator $\q_r^2$ has $M$ distinct eigenvalues 
given by $x_K^2=\frac{\hbar}{mM}(p-K)\gamma$, where $0 \leq K \leq M-1$.
The multiplicity of each $x_K^2$ is $2\binom{M-1}{K}$.  The eigenvectors
of $\q_r^2$ with eigenvalue  $x_K^2$ arise in 
two ways: there are $\binom{M-1}{K}$ vectors with $|\theta| = K+1$ of the form
\begin{equation*}
	e_{l_K0}e_{l_{K-1}0}\cdots e_{l_10}u_r;
\end{equation*}
and there are $\binom{M-1}{M-K-1}=\binom{M-1}{K}$ vectors with $|\theta| = K$ of the form
\begin{equation*}
	e_{0l_{M-K-1}} \cdots e_{0l_1} \tilde u_r. 
\end{equation*}
Herein,
\begin{equation}
	u_r = 	\sum_{k=1}^M e^{\frac{2\pi i r k}{M}} \frac{\sqrt{\beta_k}}{\omega_k} w(1^k), \qquad
  \tilde u_r = \sum_{k=1}^M e^{-\frac{2\pi i r k}{M}} (-1)^k \frac{\sqrt{\beta_k}}{\omega_k} w(0^k).
\label{u_r}	
\end{equation} 
\end{prop}

\subsection{Eigenvalues for $\q_r$}

We have shown that
the eigenvectors of $\q_r^2$ with eigenvalue $x_K^2$ have 
either $|\theta| = K$ or $|\theta| = K+1$.  Let $\psi_{r,x_K}$ be
an eigenvector of $\q_r$ with eigenvalue $x_K$.  Such an eigenvector
necessarily has the form
\begin{equation}
\psi_{r, x_K} = \sum_{|\theta| = K} C_{\theta,r, x_K} w(\theta) + 
\sum_{|\theta| = K+1} C_{\theta,r, x_K} w(\theta),
\label{psi}
\end{equation}
with $C$ some constants. Thus one can write
\begin{equation}
\sum_{|\theta| = K} C_{\theta,r, x_K} \q_r w(\theta) + 
\sum_{|\theta| = K+1} C_{\theta,r, x_K} \q_r w(\theta) = 
x_K \sum_{|\theta| = K} C_{\theta,r, x_K} w(\theta) + 
x_K \sum_{|\theta| = K+1} C_{\theta,r, x_K} w(\theta).
\label{eigv+}
\end{equation}
But the action of $\q_r$ on a basis vector $w(\theta)$ is necessarily
a linear combination of basis vectors $w(\theta')$ with $|\theta'| = |\theta|-1$ and $|\theta'| = |\theta|+1$:
this follows from~(\ref{qk}), (\ref{Qa}) and~(\ref{actiona-})-(\ref{actiona+}).
Thus the first sum on the left hand side of~(\ref{eigv+}) is 
a linear combination of basis vectors with $|\theta| = K-1$ and $|\theta| = K+1$, while the second
sum on the left hand side of~(\ref{eigv+}) is a linear combination of basis 
vectors with $|\theta| = K$ and $|\theta| = K+2$.  Of these four 
linear combinations the first and the last vanish (since they do no occur
on the right hand side), and it follows that 
\begin{equation}
\sum_{|\theta| = K} C_{\theta,r, x_K} \q_r w(\theta) = x_K \sum_{|\theta| = K+1} C_{\theta,r, x_K} w(\theta).
\label{part1-eigv+}
\end{equation}
Combining this with~(\ref{eigv+}) implies
\begin{equation}
\sum_{|\theta| = K+1} C_{\theta,r, x_K} \q_r w(\theta) = 
x_K \sum_{|\theta| = K} C_{\theta,r, x_K} w(\theta).
\label{part2-eigv+}
\end{equation}

We will now show that, given~(\ref{psi}), 
\begin{equation}
	\psi_{r,-x_K} \equiv \sum_{|\theta| = K} C_{\theta,r, x_K} w(\theta) 
	- \sum_{|\theta| = K+1} C_{\theta,r, x_K} w(\theta)
\label{phi}	
\end{equation}
is an eigenvector of $\q_r$ with eigenvalue $-x_K$.  In fact, the action of $\q_r$ on $\psi_{r,-x_K}$
follows directly from~(\ref{part1-eigv+}) and~(\ref{part2-eigv+}) and yields 
$\q_r\psi_{r,-x_K} = -x_K \psi_{r,-x_K}$.

Thus we have shown:
\begin{prop}
The operator $\q_r$ has $2M$ distinct eigenvalues 
given by $\pm x_K=\pm \sqrt{\frac{\hbar}{mM}(p-K)\gamma}$, where $0 \leq K \leq M-1$.
The multiplicity of the eigenvalue $\pm x_K$ is $\binom{M-1}{K}$.  
The eigenvectors of $\q_r$ for the eigenvalue $\pm x_K$ contain, when expanded
in the standard basis $w(\theta)$, only vectors with $|\theta|=K$ or $|\theta|=K+1$.
\end{prop}

So far, we have no simple analytic expression of the (orthogonal) eigenvectors of $\q_r$
in terms of the standard basis vectors $w(\theta)$.

\subsection{Position probability distributions for stationary states $w(\theta)$}

Consider the eigenvectors $\psi_{r,x,g}$ of the position operator $\q_r$ for the
eigenvalue $x$ expanded in the $w(\theta)$-basis:
\begin{equation}
\psi_{r,x,g} = \sum_\theta C_{\theta,r,x,g} w(\theta),
\end{equation}
and assume that these vectors are orthonormal, i.e.
\begin{equation}
\langle \psi_{r,x,g} | \psi_{r,y,h} \rangle = \sum_\theta C^*_{\theta,r,x,g}
C_{\theta,r,y,h} = \delta_{x,y}\delta_{g,h}.
\end{equation}
In the above, $g$ (or $h$) stands for a multiplicity label for vectors with
the same eigenvalue~$x$: e.g.\ when $x=\pm x_K$ then $g$
runs from 1 to $\binom{M-1}{K}$. 
We know from the previous subsection that for $x=\pm x_K$, only coefficients $C_{\theta,r,x,g}$
with $|\theta|=K$ or $|\theta|=K+1$ will appear.

Let us now suppose that the quantum system is in a fixed eigenstate $w(\theta)$ of $\hat H$
(a stationary state).
The expression
\begin{equation}
P(\theta, r, \pm x_K) = \sum_{g=1}^{\binom{M-1}{K}} | C_{\theta,r,\pm x_K,g} |^2
\label{probP}
\end{equation}
yields the probability of ``measuring'' the value $\pm x_K$ for the position of the $r$th oscillator
when the system is in the state $w(\theta)$.
Plotting all these values $P(\theta,r,\pm x_K)$ for $K=0,\ldots,M-1$ yields the probability distribution
of oscillator $r$ in the stationary state $w(\theta)$.

It will be interesting to look at an explicit example of such probability distributions.
First of all, due to the earlier mentioned symmetry of the system, these probability
distributions will be independent of~$r$; so we need to plot it for one $r$-value only (say $r=1$).
We have considered the example $M=4$, with $\hbar=m=\omega=1$ and $c=0.5$. 
Then the 8 eigenvalues $\pm x_K$ are given by $\pm \frac{\sqrt{\gamma}}{2} \sqrt{p-K}$, $K=0,1,2,3$,
with $p>3$ and $\gamma= (5\sqrt{2}+4\sqrt{3}-2)/9$ follows from~(\ref{gamma}). 
Let us also choose a value for the representation label~$p$: $p=M=4$. 
In Figure~3 we give the position probability distributions for a number of stationary states $w(\theta)$,
namely for $\theta = (0,0,0,0)$, $(1,0,0,0)$, $(0,1,0,0)$, $(0,0,1,0)$, 
$(0,0,0,1)$, $(1,1,0,0)$, $(1,1,1,0)$ and $(1,1,1,1)$. 

\begin{figure}[htb]
\begin{center}
 \begin{tabular}{ccc}
 $\theta=(0,0,0,0)$ & \ \ \ \ & $\theta=(1,0,0,0)$ \\
 \includegraphics[height=4cm,width=6cm]{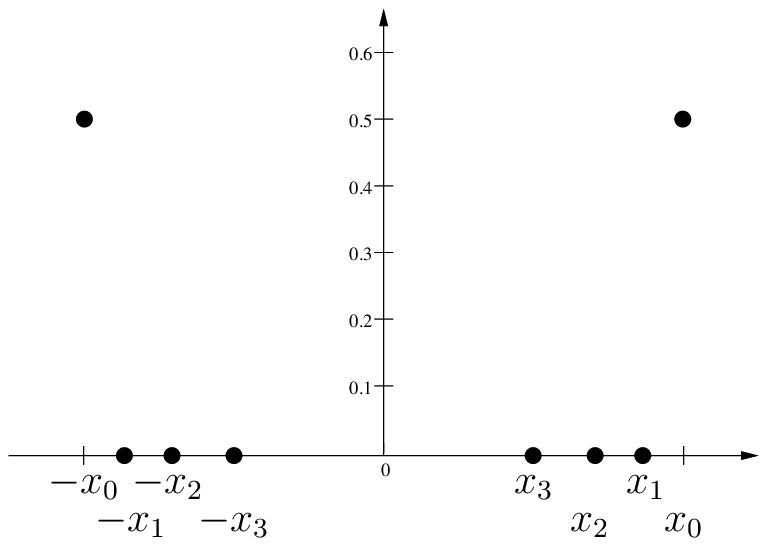} && \includegraphics[height=4cm,width=6cm]{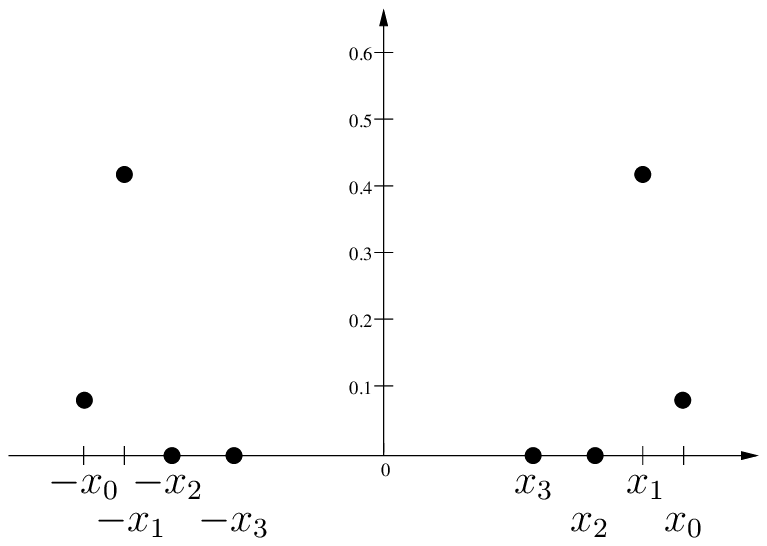}\\[8mm]
 $\theta=(0,1,0,0)$ & \ \ \ \ & $\theta=(0,0,1,0)$ \\
 \includegraphics[height=4cm,width=6cm]{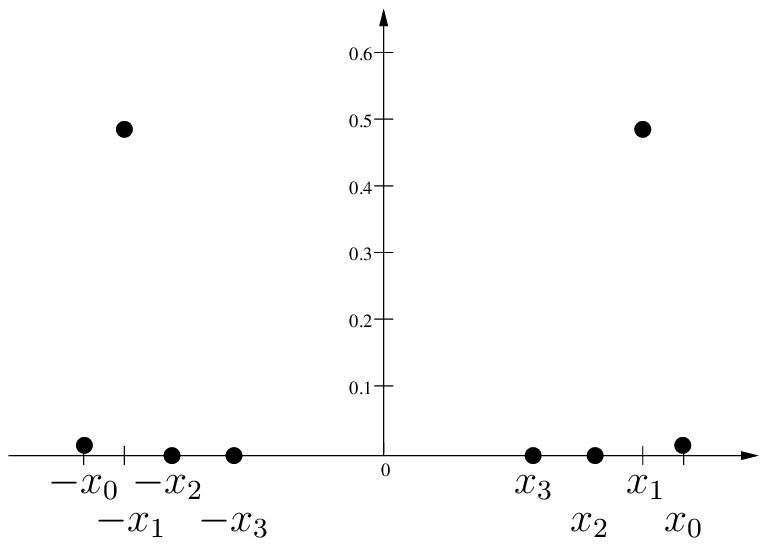} && \includegraphics[height=4cm,width=6cm]{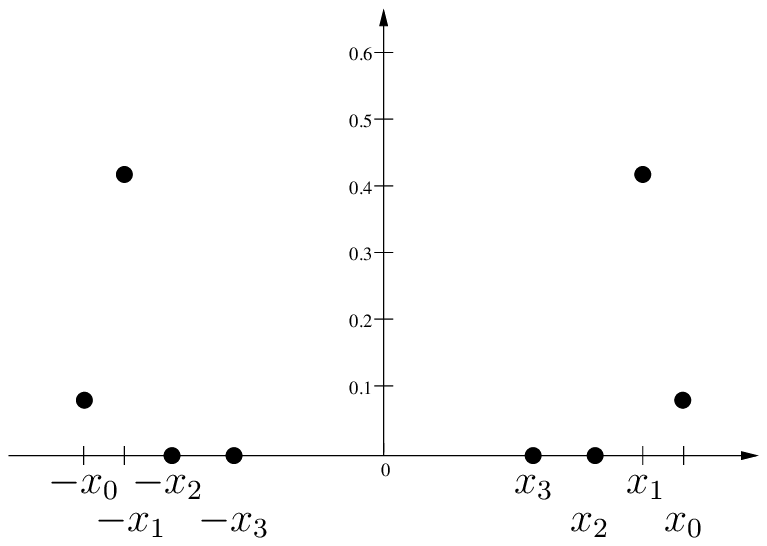}\\[8mm]
 $\theta=(0,0,0,1)$ & \ \ \ \ & $\theta=(1,1,0,0)$ \\
 \includegraphics[height=4cm,width=6cm]{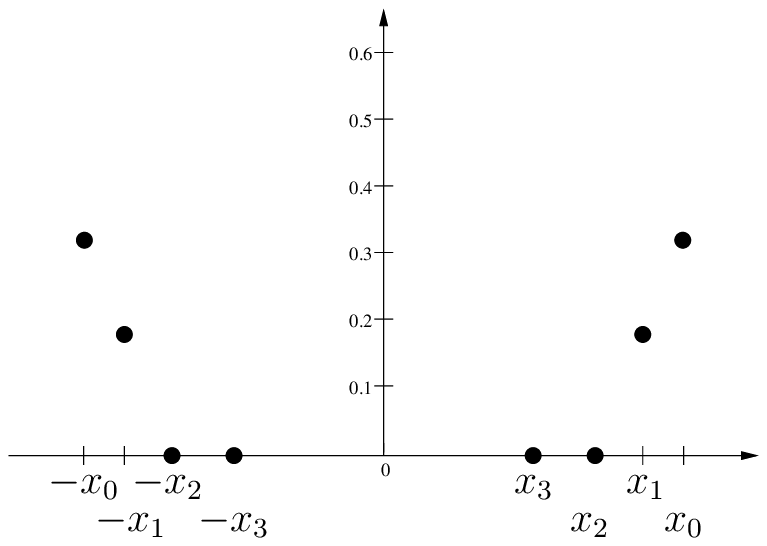} && \includegraphics[height=4cm,width=6cm]{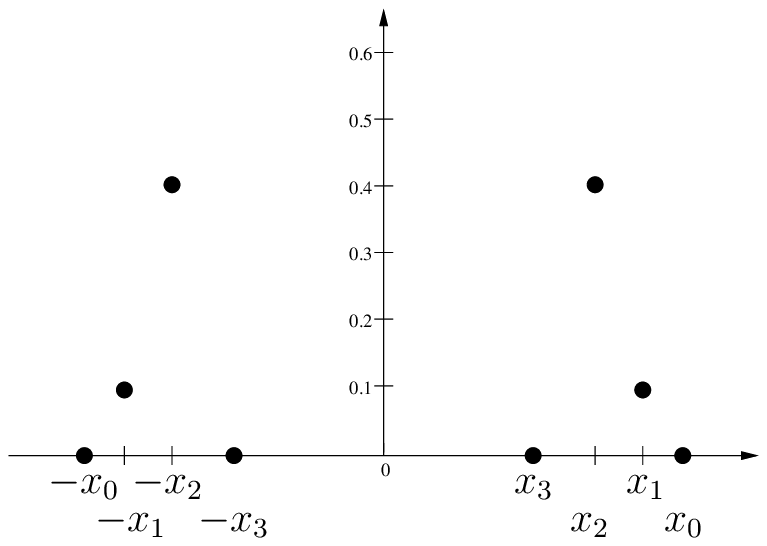}\\[8mm]
 $\theta=(1,1,1,0)$ & \ \ \ \ & $\theta=(1,1,1,1)$ \\
 \includegraphics[height=4cm,width=6cm]{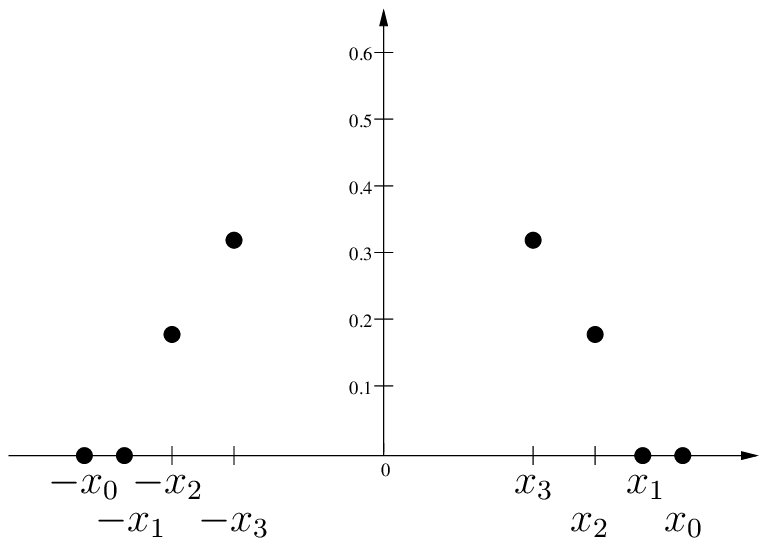} && \includegraphics[height=4cm,width=6cm]{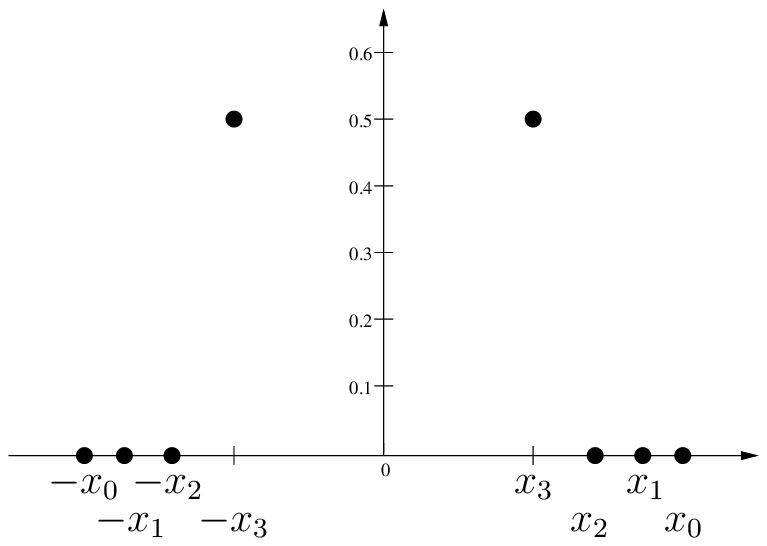}
 \end{tabular}
\end{center} 
\caption{Position probability distributions $P(\theta, 1, \pm x_K)$ for a number of $\theta$-values,
for $M=4$. The vertical axis gives the values of $P(\theta, 1, \pm x_K)$. On the horizontal
axis one finds the possible eigenvalues $\pm x_K$ with $K=0,1,2,3$.
The values of the parameters are described in the text.}
\end{figure}

Note in these plots that -- in agreement with the previous paragraph -- only the probabilities for $\pm x_K$
with $K=|\theta|$ or $K=|\theta|-1$ are nonzero.
For $\theta=(0,0,0,0)$, corresponding to the highest energy state, the oscillators can
be detected only in positions corresponding to the largest eigenvalues $x_0$ or $-x_0$. 
For $\theta=(1,1,1,1)$, corresponding to the lowest energy state, the oscillators can
be detected only in positions corresponding to the smallest eigenvalues $x_3$ or $-x_3$. 
The four plots with $|\theta|=1$ give some probability distributions in which $\pm x_0$ and $\pm x_1$
are involved. Note that for $w(1,0,0,0)$ and $w(0,0,1,0)$, two stationary states for which
the energy level is the same by~(\ref{Etheta}) and~(\ref{symm-theta}), also the probability
distributions coincide. According to~(\ref{Etheta}), 
$E_{(0,0,0,1)}> E_{(0,0,1,0)}= E_{(0,1,0,0)}> E_{(1,0,0,0)}$.
For the highest of these three energy levels, the probability of detecting the oscillator in $\pm x_0$ is larger
than detecting it in $\pm x_1$; for the lowest of these three levels, it is vice versa.
As $|\theta|$ increases (thus $E_\theta$ decreases), the probabilities indicate that
the oscillator deviation from its equilibrium position also decreases. 

\subsection{Coupling of position probability distributions}

In the previous subsection we considered, for a fixed stationary state $w(\theta)$, 
the position probabilities of the $r$th oscillator. Due to the symmetry of the system,
these probability distributions are independent of~$r$.
It will be interesting to approach the position probabilities from a different point of view.
For this purpose, let us assume that the system is in a fixed eigenstate of $\q_1$ with eigenvalue
$x$, say $\psi_{1,x,g}$. 
Let us also consider another oscillator $r\ne 1$, and the expansion of $\psi_{1,x,g}$ in terms
of the eigenvectors of $\q_r$:
\begin{equation}
\psi_{1,x,g} = \sum_{y,h} A_{1,x,g}^{r,y,h} \psi_{r,y,h}.
\end{equation}
Then
\begin{equation}
\sum_h |A_{1,x,g}^{r,y,h}|^2
\label{probA}
\end{equation}
is the probability of detecting the $r$th oscillator in the position $y$ (corresponding to
the eigenvalue $y$ of $\q_r$) when the first oscillator is in the state $\psi_{1,x,g}$.
Averaging this out over the multiplicities $g$ (if present), thus yields the probability of
detecting the $r$th oscillator in position $y$ when the first oscillator is in position $x$.

Let us again look at an example of such probability distributions.
We shall consider the same data as before: $M=4$, $\hbar=m=\omega=1$ and $c=0.5$. 
First, assume that oscillator~1 is in its highest possible position $+x_0$,
so the system is in the state $\psi_{1,+x_0,1}$. 
Then, we can compute the probabilities (\ref{probA}), for $r=2,3,4$ and for $y=\pm x_K$
($K=0,1,2,3$). These probabilities are plotted in Figure~4(a).

\begin{figure}[htb]
\begin{center}
 \begin{tabular}{ccc}
 (a) & \ \ \ \ & (b) \\[6mm]
 \includegraphics[height=34mm,width=172mm,angle=90]{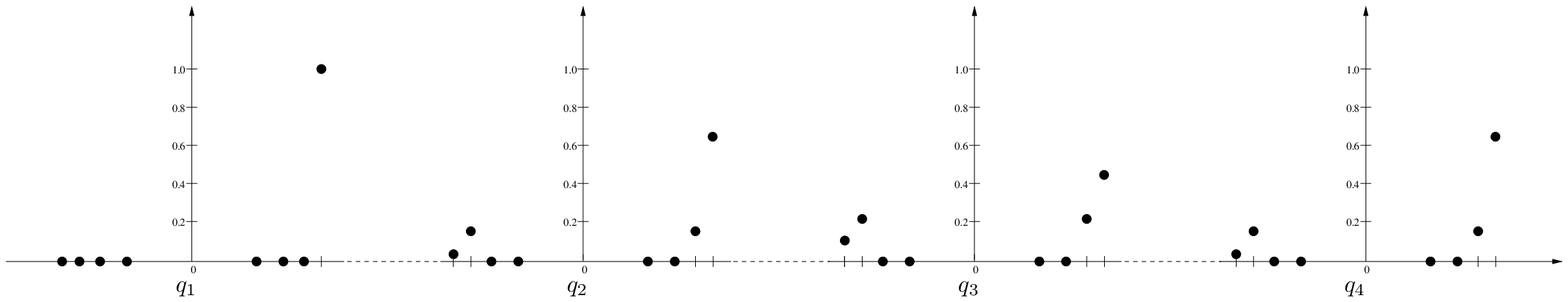} &&      
  \includegraphics[height=34mm,width=172mm,angle=90]{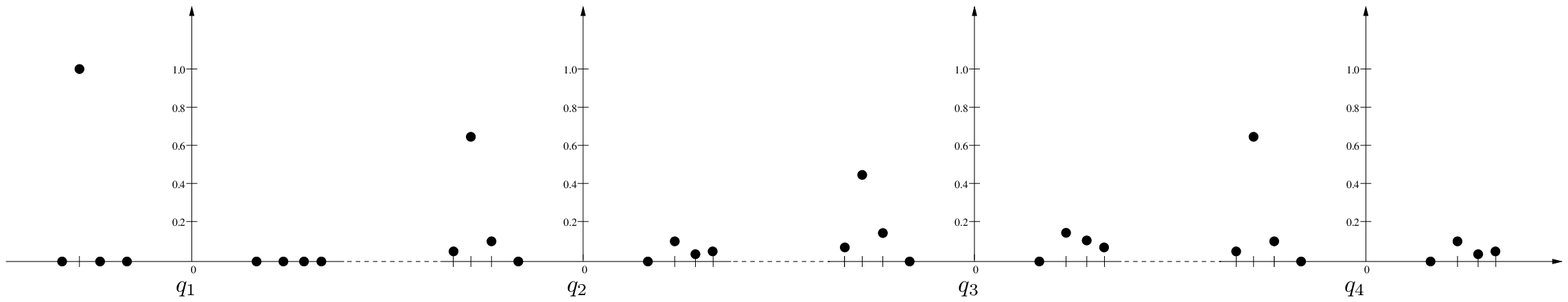}
 \end{tabular}
\end{center} 
\caption{Position probability distributions for all oscillators when the system is
in the state (a) $\psi_{1,+x_0,1}$ and (b) $\psi_{1,-x_1,g}$.}
\end{figure}

Note that the extreme position of oscillator~1 has a strong influence on the possible positions
of oscillator~2, a weaker influence on the possible positions of oscillator~3, and again a
stronger one on those of oscillator~4 (this last is due to the periodic boundary conditions~(\ref{qM+1}),
oscillator~4 behaves as if it is just to the left of oscillator~1).

We have also considered a second example, when the first oscillator is in an eigenstate
with eigenvalue $-x_1$, i.e.\ the system is in a state $\psi_{1,-x_1,g}$.
The position probability distributions for the other oscillators $r=2,3,4$ are
plotted in Figure~4(b), where we have averaged out over $g$ (here, $g=1,2,3$).
Thus in Figure~4(b) an answer is given to the following question:
suppose we make a measurement of the position of the first oscillator, and detect it
in $-x_1$, what are in that case the probabilities of finding the other oscillators 2, 3, and 4
in one of their positions $\pm x_K$?

\setcounter{equation}{0}
\section{On the spectrum of $\hat H$ and position operators in atypical 
representations $W(p)$} \label{sec:atypical}

In this section, we very briefly discuss what happens when working with
atypical representations $W(p)$, i.e.~when $p\in \{0,1,\ldots,M-1\}$.
We concentrate on the spectrum of the Hamiltonian $\hat H$ and on that of 
the operators $\hat q_r^2$ and $\hat q_r$.  Recall that the
representation space in the atypical case is a truncation of that 
in the typical case, discarding basis vectors $w(\theta)$ for
which $|\theta| > p$.  Thus one expects a close connection between
the spectra of the different operators in the typical and atypical case.

It is clear that each basis vector $w(\theta)$ is still an eigenvector of $\hat H$, with
eigenvalue $\hbar E_\theta$, with $E_\theta$ given by~\eqref{Etheta}.
So the spectrum of $\hat H$ in the atypical case is nothing but a truncation
of that in the typical case where the higher eigenvalues
are retained (of course, the actual values are different
because of the different value for $p$).
As an example, for $M=4$ and $p=1$ the dimension of the representation
space is
\begin{equation*}
\binom{4}{0} + \binom{4}{1} = 5,
\end{equation*}
and the only four eigenvalues $\hbar E_\theta$ of $\hat H$ are
\begin{align*}
&E_{(0,0,0,0)}=\beta=\beta_1+\beta_2+\beta_3+\beta_4,\quad 
E_{(0,0,0,1)}=\beta_4,\\
&E_{(1,0,0,0)}=\beta_1 = E_{(0,0,1,0)}=\beta_3,\quad
\text{and}\quad E_{(0,1,0,0)}=\beta_2,
\end{align*}
where $\beta_4>\beta_3=\beta_1>\beta_2$ for $0<c<c_0$.
These are the four topmost energy levels depicted in Figure~2(a).

In the typical case $\hat q_r^2$ has eigenvalues $x_K^2 =
\frac{\hbar}{mM}(p-K)\gamma$, with $0\leq K\leq M-1$.  {}From
equations~\eqref{qr2}, \eqref{e00}, \eqref{ejk} and~\eqref{ekj} it follows
immediately that an eigenvector of $\hat q_r^2$ in the typical case is also an
eigenvector of $\hat q_r^2$ in the atypical case provided that it is a linear
combination of basis vectors $w(\theta)$ with $|\theta| \leq p$.  When $K < p$
one has the same set of eigenvectors as in the typical case, arising from both
$u_r$ and $\tilde u_r$ and the multiplicity of $x_K^2$ is $2\binom{M-1}{K}$.
However, when $K=p$, $x_p = 0$ and only the vectors arising from $\tilde u_r$
remain (the vectors arising from $u_r$ would have $|\theta| = p+1$ which 
is impossible in an atypical representation).  
The multiplicity of eigenvalue $x_p = 0$ is thus $\binom{M-1}{p}$.
Consider the case $M=4$ and $p=1$; each operator $\hat q_r^2$ has two 
eigenvalues namely $0$, with multiplicity $\binom{4-1}{1} = 3$ and
$\frac{\hbar}{4m}(1)\gamma$ with multiplicity $2\binom{3}{0} = 2$.

For the position operators $\hat q_r$ finally, it is seen as before
that $\pm x_K$ with $0\leq K \leq p-1$ are eigenvalues of $\hat q_r$
each with multiplicity $\binom{M-1}{K}$.   Besides these eigenvalues,
there is also the eigenvalue $x_p = 0$ with multiplicity $\binom{M-1}{p}$.
So in the atypical case it is possible to \lq\lq detect\rq\rq\ an
oscillator in its equilibrium position, in contrast with the 
typical case.

It is worth giving some further details for the representation $W(1)$
(so $p=1$), for general $M$-values. This representation has dimension $M+1$,
with basis vectors $w(0)$ and $w(1^j)$ (in the notation of~(\ref{v_r})), with
$j=1,\ldots,M$. 
Each position operator $\q_r$ has spectrum $\{-x_0, 0, +x_0\}$, with
multiplicities $\{1,M-1,1\}$ respectively, where $x_0=\sqrt{\frac{\hbar\gamma}{mM} }$.
Herein, $\gamma$ is given by~(\ref{gamma}); in fact it will be useful to
introduce the notation
\begin{equation}
\gamma_k = \frac{\sqrt{\beta_k}}{\omega_k},\qquad k=1,\ldots,M,
\label{gamma_k}
\end{equation}
and thus $\gamma=\sum_{k=1}^M \gamma_k^2$.

In this case, it is not difficult to construct explicitly a set
of orthonormal eigenvectors of $\q_r$. In the notation of subsection~6.3, it is given by:
\begin{align}
\psi_{r,+x_0,1} & = \frac{1}{\sqrt{2}}\; w(0) + \sum_{j=1}^M \frac{\gamma_j}{\sqrt{2\gamma}}\,
e^{\frac{2\pi i rj}{M}} w(1^j),\\
\psi_{r,-x_0,1} & = \frac{1}{\sqrt{2}}\; w(0) - \sum_{j=1}^M \frac{\gamma_j}{\sqrt{2\gamma}}\,
e^{\frac{2\pi i rj}{M}} w(1^j),\\
\psi_{r,0,g} & = \frac{1}{\sqrt{ \frac{1}{\gamma_1^2+\cdots+\gamma_g^2}+\frac{1}{\gamma_{g+1}^2} }}
\left(\sum_{j=1}^g \frac{\gamma_j}{\gamma_1^2+\cdots+\gamma_g^2 }\,
e^{\frac{2\pi i rj}{M}} w(1^j) - \frac{1}{\gamma_{g+1}}\, e^{\frac{2\pi i r(g+1)}{M}} w(1^{g+1})\right),
\end{align}
where $g=1,2,\ldots,M-1$.

Now it becomes simple to compute some position probabilities.
Following~(\ref{probP}), one finds for the state $w(0)$ with $\theta=(0)=(0,\ldots,0)$:
\begin{equation*}
P((0),r,+x_0) = P((0),r,-x_0) = \frac{1}{2}, \quad P((0),r,0) = 0.
\end{equation*}
Hence in the highest energy state of $W(1)$, the oscillators can be detected only
in the positions $\pm x_0$ and not in~0.
Similarly:
\begin{equation*}
P((1^j),r,+x_0) = P((1^j),r,-x_0) = \frac{\gamma_j^2}{2\gamma}, \quad 
P((1^j),r,0) = 1-\frac{\gamma_j^2}{\gamma} .
\end{equation*}

Also the probabilities~(\ref{probA}) can be computed. 
One finds, for example:
\begin{equation}
A_{1,+x_0,1}^{r,+x_0,1} = \frac{1}{\gamma} \sum_{j=1}^M \gamma_j^2 \cos^2(\frac{\pi(r-1)j}{M}),\qquad
A_{1,+x_0,1}^{r,-x_0,1} = \frac{1}{\gamma} \sum_{j=1}^M \gamma_j^2 \sin^2(\frac{\pi(r-1)j}{M}),
\label{cossin}
\end{equation}
and thus
\begin{equation*}
\sum_g |A_{1,+x_0,1}^{r,0,g}|^2 = \frac{2}{\gamma^2} \left(\sum_{j=1}^M \gamma_j^2 \cos^2(\frac{\pi(r-1)j}{M})\right) \left(\sum_{j=1}^M \gamma_j^2 \sin^2(\frac{\pi(r-1)j}{M})\right).
\end{equation*}
As before, the quantities $|A_{1,+x_0,1}^{r,+x_0,1}|^2$, $\sum_g |A_{1,+x_0,1}^{r,0,g}|^2$ and
$|A_{1,+x_0,1}^{r,-x_0,1}|^2$ describe the probabilities of detecting the $r$th oscillator in
the position $+x_0$, $0$ or $-x_0$ respectively, when the first oscillator is in its highest position~$+x_0$.

With the given probabilities, one can consider a final illustration. 
If the system is in the state $\psi_{1,+x_0,1}$,
the average position of each oscillator~$r$ is given by
\begin{equation*}
+x_0 |A_{1,+x_0,1}^{r,+x_0,1}|^2 + 0 \sum_g |A_{1,+x_0,1}^{r,0,g}|^2 - x_0 |A_{1,+x_0,1}^{r,-x_0,1}|^2.
\end{equation*}
Using~(\ref{cossin}), this simplifies to
\begin{equation*}
\langle \q_r \rangle _{\psi_{1,+x_0,1}} = 
x_0 \frac{1}{\gamma} \sum_{j=1}^M \gamma_j^2 \cos(\frac{2\pi(r-1)j}{M}).
\end{equation*}
In Figure~5 we plot the average position of each oscillator~$r$ in this state.
So in this figure one can see the effect of having the first oscillator in its
highest position $+x_0$ on the average position of the other oscillators.
\begin{figure}[htb]
\begin{center}
  \includegraphics{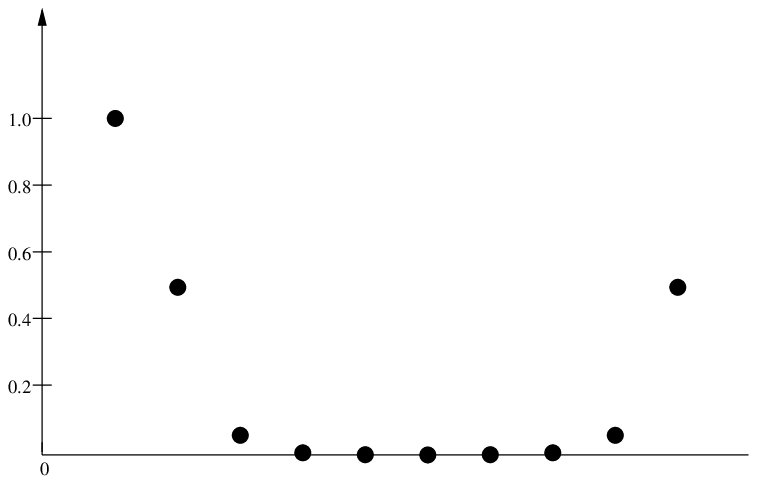}
\end{center} 
\caption{Average positions of the $M$ oscillators, $M=10$, when the first oscillator
is in the position $+x_0$, in the representation $W(1)$.
Here, we have taken $c=0.12\,\omega^2$.
The horizontal axis labels the $M$ oscillators; the vertical axis gives the average
position in units of $x_0$.}
\end{figure}

\setcounter{equation}{0}
\section{Conclusion and outlook} \label{sec:conclusion}

We have examined some properties of non-canonical solutions of the quantum
system determined by the Hamiltonian~(\ref{Intro-H}).
These solutions arise from a Wigner Quantum Systems approach, 
where the quantization conditions are weaker than the canonical commutation relations,
thus allowing more types of solutions.

The solutions studied here are found by identifying certain linear combinations
of position and momentum operators with generators of the Lie superalgebra $\gl(1|M)$.
We have shown that this is always possible, but that the solutions are corresponding to
the compact form $\u(1|M)$ of $\gl(1|M)$ only if the coupling constant $c$ is sufficiently small.

The physical Hilbert spaces in which the states of the system are described then
correspond to unitary representations of $\gl(1|M)$. In this paper, we have considered
only a simple class of such unitary representations, the so-called Fock spaces $W(p)$.
This class of representations turns out to be already sufficiently resourceful in order
to exhibit some fascinating physical properties of the solutions. 
Of special interest is the feature of having only a discrete spectrum for each
oscillator position operator. At first sight, this is somewhat unusual.
On the other hand, our analysis of probability distributions for position operators
has shown effects that are reminiscent of canonical results. 

This paper presents only the first results for this quantum system consisting of a
one-dimensional chain of coupled harmonic oscillators in the Wigner Quantum Systems approach.
There are still many open problems or new aspects to be studied.
For example, it is clear that the system~(\ref{algrelations}) has also solutions 
outside $\gl(1|M)$. For instance, if $M=2n$ is even, then one can construct 
an algebraic solution by means of the direct sum Lie superalgebra $\gl(1|2) \oplus
\cdots \oplus \gl(1|2)$ ($n$ copies). In this case, the unitarity conditions
following from the form $\u(1|2) \oplus \cdots \oplus \u(1|2)$ imply no conditions
on the coupling constant $c$, and it would be interesting to study the system from this point of view.
Furthermore, it would be worth investigating whether~(\ref{algrelations}) has also
solutions related to orthosymplectic Lie superalgebras~\cite{SV2}.

But even if we restrict for the moment our attention to the $\gl(1|M)$ solutions
given here in Section~\ref{sec:alg}, work remains to be done. In particular, 
one should also consider other classes of unitary $\gl(1|M)$ representations~\cite{KSV}
and investigate the corresponding physical properties.

Finally, and this a more technical question, the explicit construction of orthonormal
eigenvectors of the position operators $\q_k$ is lacking.
Although this can be done numerically for any given $M$ and a given set
of parameters, we have at the moment no closed form expressions for these eigenvectors.
We hope to find such forms, as they would allow us to draw some general
conclusions regarding position probability distributions.
At this moment, the last conclusions in Section~6 are based upon
observations of examples rather than upon analytic formulas.
We expect to return to some of these remaining questions in future publications. 

\section*{Acknowledgments}
The authors would like to thank Professor T.D.~Palev for his interest.
NIS was supported by a project from the Fund for Scientific Research -- Flanders (Belgium).


\begin{thebibliography}{99}
\bibitem{Audenaert}
Audenaert K, Eisert J, Plenio M B and Werner R F 2002 {\it Phys Rev A } {\bf 66} 042327
\bibitem{Brun} 
Brun T A and Hartle J B 1999 {\it Phys Rev D } {\bf 60} 123503
\bibitem{Eisert}
Eisert J and Plenio M B 2003 {\it Int. J. Quant. Inf.} {\bf 1} 479
\bibitem{Halliwell}
Halliwell J J 2003 {\it Phys Rev D} {\bf 68} 025018
\bibitem{Plenio}
Plenio M B, Hartley J and Eisert J 2004 {\it New J. Phys.} {\bf 6} 36
\bibitem{Cohen}
Cohen-Tannoudji C, Diu B and Lalo\"e F 1977 {\em Quantum Mechanics}, John Wiley and Sons, New York
[volume~1, complement JV]
\bibitem{Palev86}
Kamupingene A H, Palev T D and Tsavena S P 1986 {\em J.\ Math.\ Phys.} {\bf 27} 2067
\bibitem{Wigner}
Wigner E P 1950 {\it Phys. Rev.} {\bf 77} 711
\bibitem{Green}
Green H S 1953 {\em Phys.\ Rev.} {\bf 90} 270 
\bibitem{Kamefuchi}
Kamefuchi S and Takahashi Y 1962 {\em Nucl.\ Phys.} {\bf 36} 177
\bibitem{Ryan}
Ryan C and Sudarshan E C G 1963 {\em Nucl.\ Phys.} {\bf 47} 207 
\bibitem{Palev1}
Palev T D 1982 {\it J.\ Math.\ Phys.} {\bf 23} 1778; 1982 {\it Czech.\ J.\ Phys.} 
\bibitem{Degasperis}
Degasperis A and Ruijsenaars S N M 2001 {\it Ann. Phys.} {\bf 293} 92
\bibitem{Atakishiyev}
Atakishiyev N M, Pogosyan G S and Wolf K B 2005 {\it Phys. Part. Nuclei} {\bf 36} 247
\bibitem{Lohe}
Lohe M A 2006 {\it Rep. Math. Phys.} {\bf 57} 131
\bibitem{PS1}
Palev T D and Stoilova N I 1997 {\it J.\ Math.\ Phys.} {\bf 38} 2506
\bibitem{PS2}
Palev T D and Stoilova N I 1994 {\it J. Phys. A: Math. Gen.} {\bf 27} 7387
{\bf B32} 680
\bibitem{KPSV1}
King R C, Palev T D, Stoilova N I and Van der Jeugt J 2003 {\it J. Phys. A: Math. Gen.} {\bf 36} 4337
\bibitem{KPSV2}
King R C, Palev T D, Stoilova N I and Van der Jeugt J 2003 {\it J. Phys. A: Math. Gen.} {\bf 36} 11999
\bibitem{Palev2}
Palev T D 2006 {\it $SL(3|N)$ Wigner quantum oscillators: examples of ferromagnetic-like
oscillators with noncommutative, square-commutative geometry}, preprint hep-th/0601201
\bibitem{Kac1}
Kac V G 1977 {\it Adv.\ Math.} {\bf 26} 8
\bibitem{Kac2}
Kac V G 1978 {\it Lect.\ Notes Math.} {\bf 676} 597 
\bibitem{Scheunert}
Scheunert M 1979 {\em The theory of Lie superalgebras: an introduction}, 
Lectures Notes in Mathematics {\bf 716}
\bibitem{Parker}
Parker M  1980 {\em J. Math. Phys.} {\bf 21} 689 
\bibitem{Gould} 
Gould M D and Zhang R B  1990 {\it J.\ Math.\ Phys.} {\bf 31} 2552 
\bibitem{KSV}
King R C, Stoilova N I and Van der Jeugt J 2006 {\it J. Phys. A: Math. Gen.} {\bf 39} 5763
\bibitem{Palev0}
Palev T D 1980 {\it J.\ Math.\ Phys.} {\bf 21} 1293
\bibitem{SV1}
Stoilova N I and Van der Jeugt J 2005 {\it Int. J. Theor. Phys.} {\bf 44} 1157
\bibitem{JHKT}
Van der Jeugt J, Hughes J W B, King R C and Thierry-Mieg J 1990
{\it J\ Math\ Phys} {\bf 31} 2278
\bibitem{Sloane}
Sloane N J A 1996-2006 {\em The on-line encyclopedia of integer sequences}, published electronically at http://www.research.att.com/njas/sequences/
\bibitem{SV2}
Stoilova N I and Van der Jeugt J 2005 {\it J. Phys. A: Math. Gen.} {\bf 38} 9681


\end{thebibliography}
\end{document}